\documentclass[twocolumn]{openjournal}
\usepackage{natbib}
\usepackage{graphicx}
\usepackage{url} 
\newcommand*{\rdash}{--}   
\newcommand*{\jdash}{--}   
\newcommand*{\ndash}{--}   
\newcommand*{\kdash}{ -- } 
\newcommand*{\pdash}{---}  
\newcommand*{\tdash}{---}  
\newcommand*{\hnull}{\ensuremath{H_{0}}}

\newcommand*{\lnull}{\ensuremath{\lambda_{0}}}
\newcommand*{\onull}{\ensuremath{\Omega_{0}}}
\newcommand*{\qnull}{\ensuremath{q_{0}}}
\newcommand*{\snull}{\ensuremath{\sigma_{0}}}

\newcommand*{\D}{\mathop{}\!\mathrm{d}}

\newcommand*{\plusminus}[2]{^{+#1}_{-#2}}
\newcommand*{\unit}{~}
\newcommand*{\mathspace}{\quad}
\providecommand*{\degree}{^{\circ}}
\newcommand*{\asd}{angular-size distance}
\newcommand*{\asrr}{angular-size\jdash{}redshift relation}
\newcommand*{\Dasd}{\ensuremath{D^{\mathrm{A}}}}
\newcommand*{\Dd}{\ensuremath{D_{\mathrm{d}}}}
\newcommand*{\Dds}{\ensuremath{D_{\mathrm{ds}}}}
\newcommand*{\Dl}{\ensuremath{D^{\mathrm{L}}}}
\newcommand*{\Dp}{\ensuremath{D^{\mathrm{P}}}}

\newcommand*{\Ds}{\ensuremath{D_{\mathrm{s}}}}
\newcommand*{\EdS}{Einstein\ndash{}de~Sitter}
\newcommand*{\flrw}{Friedmann\ndash{}Lema\^{\i}tre\ndash{}Robertson\ndash{}Walker}

\newcommand*{\flrwfirst}{\flrw\ (FLRW)}
\newcommand*{\frw}{Friedmann\ndash{}Robertson\ndash{}Walker}
\newcommand*{\FRW}{FRW}
\newcommand*{\aFRW}{an FRW}
\newcommand*{\frwfirst}{\frw\ (FRW)}
\newcommand*{\lcdm}{$\Lambda$CDM}

\newcommand*{\LTB}{Lema\^{\i}tre\ndash{}Tolman\ndash{}Bondi}
\newcommand*{\mzr}{$m$\jdash{}$z$ relation}
\newcommand*{\rdr}{redshift\jdash{}distance relation}

\newcommand*{\tIas}{Type Ia supernovae}
\newcommand*{\zd}{\ensuremath{z_\mathrm{d}}}
\newcommand*{\zl}{\ensuremath{z_\mathrm{l}}}
\newcommand*{\zs}{\ensuremath{z_\mathrm{s}}}
\newcommand*{\foreign}[1]{\emph{#1}}
\newcommand*{\adhoc}{\foreign{ad~hoc}}
\newcommand*{\ansatz}{\foreign{ansatz}}
\newcommand*{\apriori}{\foreign{a~priori}}
\newcommand*{\cf}{\foreign{cf.}}

\newcommand*{\etc}{\foreign{etc.}}
\newcommand*{\etcnp}{\foreign{etc}}
\newcommand*{\eg}{\foreign{e.g.}}

\newcommand*{\gedankenexperiment}{\foreign{gedankenexperiment}}
\newcommand*{\ie}{\foreign{i.e.}}

\newcommand*{\percent}{~per~cent}
\newcommand*{\percentadj}{-per-cent}
\newcommand*{\perse}{\foreign{per se}}
\newcommand*{\sic}{\foreign{sic}}
\newcommand*{\via}{\foreign{via}}
\newcommand*{\viceversa}{\foreign{vice versa}}
\newcommand*{\eq}[1]{Eq.~(\ref{#1})}
\newcommand*{\Eq}[1]{Eq.~(\ref{#1})}

\newcommand*{\eqn}[1]{equation~(#1)}
\newcommand*{\Eqn}[1]{Equation~(#1)}
\newcommand*{\eqns}[3]{equations~(#1)#2(#3)}
\newcommand*{\Eqns}[3]{Equations~(#1)#2(#3)}
\newcommand*{\fig}[1]{Fig.~\ref{#1}}
\newcommand*{\Fig}[1]{Fig.~\ref{#1}}

\newcommand*{\figr}[1]{figure~#1}
\newcommand*{\Figr}[1]{Figure~#1}

\newcommand*{\Figrs}[3]{Figures~#1#2#3}
\newcommand*{\sect}[1]{Sect.~\ref{#1}} 
\newcommand*{\Sect}[1]{Sect.~\ref{#1}}
\newcommand*{\sects}[3]{Sects.~\ref{#1}#2\ref{#3}}

\newcommand*{\tab}[1]{Tab.~\ref{#1}}

\newcommand*{\tabl}[1]{table~#1}

\newcommand*{\ack}{Acknowledgements}
\newcommand*{\bestfit}{best-fit}

\newcommand*{\haloes}{haloes}
\newcommand*{\lightyears}{light-years} 

\newcommand*{\spacetime}{spacetime}
\newcommand*{\arcminw}{arcmin} 

\newcommand*{\arcminutea}{arcminute}
\newcommand*{\subarcminutea}{sub-arcminute}
\newcommand*{\arcsecw}{arcsec}

\newcommand*{\code}[1]{{\sc #1}}
\newcommand*{\degrees}{degrees}
\newcommand*{\journal}[1]{\emph{#1}}
\newcommand*{\magnitude}{mag} 
\newcommand*{\satellite}[1]{\emph{#1}}
\newcommand*{\software}[1]{{\sc #1}}
\newcommand*{\Universe}{Universe} 
\newcommand*{\universe}{universe} 
\newcommand*{\coauthored}{\mbox{(co-)authored}}
\begin{document}

\title{Calculation of distances in cosmological models with small-scale
inhomogeneities and their use in observational cosmology: a review}

\shorttitle{Distances in inhomogeneous cosmological models}

\author{Phillip Helbig}

\shortauthors{Phillip Helbig}

\email{phillip.helbig@doct.uliege.be, helbig@astro.multivax.de}

\affil{Institut d'Astrophysique et de G{\'e}ophysique (B{\^a}t B5c)\\
Universit{\'e} de Li{\`e}ge\\
Quartier Agora\\
All{\'e}e du 6 ao{\^u}t, 19C \\
B-4000 Li{\`e}ge 1 (Sart-Tilman)\\
Belgium}

\received{September 12, 2019}
\revised{November 17, 2019}
\accepted{December 23, 2019}

\begin{abstract}
\noindent 
The \Universe{} is not completely homogeneous.  Even if it is
sufficiently so on large scales, it is very inhomogeneous at small
scales, and this has an effect on light propagation, so that the
distance as a function of redshift, which in many cases is defined
\via{} light propagation, can differ from the homogeneous case.  Simple
models can take this into account.  I review the history of this idea,
its generalization to a wide variety of cosmological models, analytic
solutions of simple models, comparison of such solutions with exact
solutions and numerical simulations, applications, simpler analytic
approximations to the distance equations, and (for all of these aspects)
the related concept of a `Swiss-cheese' \universe{}. 
\end{abstract}


\maketitle

\section{Introduction}

The \Universe{} is not completely homogeneous; if it were, there would
be no observers and no objects to be observed.  Nevertheless, distances
are often calculated as a function of redshift as if that were the case,
at least as far as light propagation is concerned.  Whether this is a
good approximation depends at least on the angular scale involved.  The
simplest more refined model retains the background geometry and
expansion history of a \frwfirst{} model but separates matter into two
components, one smoothly distributed comprising the fraction $\eta$ of
the total density and the other ($1-\eta$) consisting of clumps, and
considers the case where light from a distance object propagates far
from all clumps (this is equivalent to the case of negligible shear).
Over a period of more than 50 years, various authors have described
more-general versions of this approximation with regard to cosmology,
found analytic solutions, discussed similar approximations, compared it
with exact solutions and with brute-force numerical integration based on
the gravitational deflection of matter along and near the line of sight,
examined the assumptions involved, applied it to various cosmological
and astrophysical problems, and developed simple analytic approximations
both for more-exact solutions (the latter based on more-complicated
analytic formulae or on numerical integration) and for numerical
simulations.  While there is no doubt that such an approximation is
valid for a \universe{} with the corresponding mass distribution, recent
work indicates that our \Universe{} is not such a \universe{}, but
rather one in which the `standard distance' (\ie{}~calculated under the
assumption of strict homogeneity) is valid, even for small angular
scales, at least to a good approximation. 

Further refinements of this approximation are not discussed here,
\eg{}~weak gravitational lensing with non-negligible shear, strong
gravitational lensing%
\footnote{While some cases of strong gravitational lensing are
discussed, in most cases these are not concerned with the influence of
the lensing effect on the distance; rather, the approximations discussed
are used to calculate the distances involved, the strong lensing being
calculated explicitly.%
}%
, or inhomogeneities so appreciable that they influence the large-scale
geometry and/or expansion history of the \universe{} (`back reaction').
Similarly, extensions to the \FRW{} models, such as some sort of `dark
energy' other than the cosmological constant, are not considered;
neither are those which violate the Cosmological Principle, \eg{}~ones
in which we are within a large void, \LTB{} models, \etcnp{}.  I also
omit wrong results or misleading conclusions unless they have been often
cited without all of the community noticing the mistake (either there
was no correction or the correction has been ignored). 

The order is chronological in the sense that I discuss all papers on the
first topic to appear, then all on the second topic, and so on. 

I refer to the distance calculated based on the above approximation as
the \emph{ZKDR} distance, a term introduced by 
\citet{RSantosJLima06a}
and referring to Zel'dovich, Kantowski, Dyer, and Roeder, though I take
the `D' to refer to Dashevskii as well, my criteria for being part of
the acronym being having \coauthored{} at least two papers on this
topic, at least one of which was published within ten years of the first
paper on this topic 
\citep{YZeldovich64ao,YZeldovich64at}. 

In gravitational lensing, it is clear that the approximation of a
completely homogeneous \universe{} with regard to light propagation
cannot be valid, since otherwise there would be no gravitational
lensing.  Perhaps for this reason, the ZKDR distance has been used more
in gravitational lensing than in other fields.  Since $\alpha$ is almost
universally used to denote the gravitationl-lensing bending angle,
\citet{RKayserHS97Ra}, 
hereafter KHS, adopted $\eta$ instead of the more confusing $\alpha$ or
$\tilde{\alpha}$ used by some other authors; since then, some authors
other than KHS have also used $\eta$ instead of $\alpha$ or
$\tilde{\alpha}$ for the inhomogeneity parameter.  In the following, I
will use the notation of KHS except occasionally when explicitly
referring to equations in the works of other authors, who use various
and sometimes confusing notation schemes\tdash{}in particular, using $z$
for anything other than redshift in a paper on cosmology is very
confusing (see \tab{Z64-t}).

\section{Zel'dovich (1964)}

\citet[hereafter Z64]{YZeldovich64at}%
\footnote{This discussion follows the English translation of the Russian
original 
\citep{YZeldovich64ao}.%
} 
started the tradition; many (not only) today might find his paper
somewhat idiosyncratic, difficult to follow, and wrong in parts, but he
introduced a simple and useful basic idea: local inhomogeneities in the
distribution of matter can lead to significantly different angular-size
and luminosity densities from those derived under the assumption of a
perfect \FRW{} model.

\subsection{Summary}

The first attempt to calculate distances in a \universe{} with
small-scale inhomogeneities is, as far as I know, that of Z64.  This
begins a tradition of calculating distances in a more realistic
\universe{}, namely one with small-scale inhomogeneities, but where the
large-scale dynamics is given by \aFRW{} model.  In other words, it is a
perturbed \FRW{} model:  The zeroth-order approximation for cosmology,
which is actually quite good 
\citep{SGreenRWald14a},
is that the \Universe{} is described by a Robertson\ndash{}Walker metric
\citep[][the latter paper by Walker is very often incorrectly cited as
having been published in 1936]%
{HRobertson1935a,HRobertson1936a,AWalker35b,AWalker37a} 
which is a purely descriptive kinematic idea with no physics content,
merely the characterization of a homogeneous and isotropic \universe{},
and that the expansion history is given by one of the models explored by
\citet{AFriedmann22a,AFriedmann24a}
(hence \FRW{}), which are in turn based on relativistic cosmology as
introduced by 
\citet{AEinstein17a}.
Occasionally, the term \flrwfirst{} is used to include a reference to 
\citet{GLemaitre27a};
while he made important contributions to cosmology, none of them went
beyond the work of 
\citet{AFriedmann22a,AFriedmann24a}
with respect to the metric.  `It is assumed that\dots{}the amount of
matter removed is small and the general motion is not affected.'  The
main model considered is `a flat Friedman [sic] model with pressure
equal to zero'.  In modern notation, $\onull{}=1$ and $\lnull{}=0$.  The
physical model assumes that all matter exists in galaxies%
\footnote{More precisely, that the mass of the intergalactic medium can
be neglected compared to the mass of matter contained in galaxies.%
} 
and that distant objects are seen between galaxies, \ie{}~such distant
objects `do not have galaxies within the cone subtended by them at the
observer'.  (The cone is often referred to as the beam.) 

After the standard \asd{}%
\footnote{His \eqn{7} assumes the \EdS{} model (introduced at the start
of the appendix as the `flat Friedman [\sic{}] model with pressure equal
to zero'); a casual reader might think that it applies more generally.%
}
is derived \via{} a differential equation for the separation between two
light rays, the deflection due to a point mass, \eqn{12}, is used to
calculate the deviation from the completely homogeneous case when the
beam is devoid of matter.  This equation is generalized to a uniform
density distribution to calculate the total deflection, which is towards
the outside since the removal of matter in the beam formally corresponds
to negative mass.  This leads to a differential equation which in turn
leads to the expression for the \asd{} in the \EdS{} model in the
empty-beam case, denoted by $f_{1}$ in Z64.  It is noted that this
function `increases monotonically right up to the [particle] horizon
($\Delta=1$) where it reaches the value 2/5'.  The value 2/5 is exact,
but the right-hand side of the unnumbered equation between \eqns{21}{
and }{22}, 1600, is too precise (though the correct value rounded to
four digits is 1599, much closer than in the case discussed in
\sect{Z64-remarks}). 

It is noted that `the calculation can be repeated for the case when
$\rho\neq\rho_{\mathrm{c}}$, \ie{}, for a hyperbolic or closed
universe', though this is given (without derivation) only for the
`limiting case $\rho\rightarrow 0$, Milne model'.  \Eqns{23}{ and }{24},
for the Milne and \EdS{} models respectively, are of course special
cases of the formulae derived by 
\citet{WMattig58a},
who gives a simple formula for $\onull{}=0$, essentially the same as
that of Z64, and a more complicated formula for $\onull{}>0$ (though
assuming $\lnull{}=0$); see KHS, \eqns{B24}{ and }{B25}. 
In the case of $\onull{}=0$, the value of $\eta$ doesn't matter. 
\Eqn{25}, 
\begin{displaymath}
f_{1} = 2/5[1 - (1 - \Delta)^{5/2}] \mathspace{},
\end{displaymath}
is, in modern notation,
\begin{displaymath}
\frac{\hnull{}}{c}\Dasd{} = \frac{2}{5}
     \left[1 - (z+1)^{-\frac{5}{2}}\right] 
\end{displaymath}
(\cf{}~KHS, \eqn{B15 (II)}).

\subsection{Remarks}
\label{Z64-remarks}

There are several strange things about this paper.  First, the
`remarkable feature' that the \asd{} has a maximum is noted.  Second, it
is claimed that this `is caused by the curvature of space due to the
matter filling the universe', which is strange because later in the
paper the main model considered is a \emph{flat} \universe{}, \ie{}~one
with no spatial curvature.  Third, it is pointed out that the maximum
`occurs only when there is matter within the cone subtended by the
object at the point of observation'.  Fourth, for a modern reader, the
notation is extremely bizarre; \tab{Z64-t} shows the equivalents
in modern notation of the quantities used.  I now discuss each of these
in turn. 
\begin{table}
\caption{Note that $t$, $t_{0}$, $c$, $\rho$ are the same in the Z64 and
modern notations.  Z64 distinguishes between $\Theta$ and $\Theta_{1}$
for the cases $\eta=1$ and $\eta=0$, respectively, though in both cases
the quantity is the \emph{observed} angular size.  Similarly,
$\Theta_{2}$ is the observed angular size in the case of strong
gravitational lensing.  Except in the case of $f_{0}$, quantities
dependent on the cosmological model assume the \EdS{} model.} 
\label{Z64-t}
\begin{tabular}{ll}
&\\
\hline
\hline
Z64 notation               &       modern notation                \\
\hline
$\Theta$                   &       $\theta$                       \\
$\Theta_{1}$               &       $\theta$                       \\
$\Theta_{2}$               &       $\theta$                       \\
$\Delta$                   &       $1 - (1+z)^{-1}$               \\
$(1-\Delta)^{-1} - 1$      &       $z$                            \\
$\omega_{1}$               &       $\omega_{0}$                   \\
$\omega_{0}$               &       $\omega$                       \\
$r$                        &       $l$ or $\ell$                  \\
$f$                        &       $Hc^{-1}\Dasd{}$ ($\eta=1$)    \\
$f_{1}$                    &       $Hc^{-1}\Dasd{}$ ($\eta=0$)    \\
$f_{0}$                    &       $Hc^{-1}\Dasd{}$ (Milne model) \\
$\kappa$                   &       $G$                            \\
$H$                        &       $H_{0}$                        \\
$R$                        &       \Dasd{} ($\eta=1$)             \\
$R_1$                      &       \Dasd{} ($\eta=0$)             \\
Mps                        &       Mpc                            \\
\hline
\vspace{1ex}
\end{tabular}           
\end{table}

What is remarkable about the fact that the \asd{} has a maximum at some
redshift?  In modern notation, the \asd{} \Dasd{} is, by definition,
$l/\theta$, where $l$ is the physical projected length of the object and
$\theta$ the angle which it subtends, \ie{}~the angle at the observer
formed by light rays from both ends of the object.%
\footnote{See the paper by KHS for definitions of various cosmological
distances which are consistent, use modern notation, and deviate as
little as possible from the approximate consensus.%
}
The triangle made by the object and the light rays retains its shape as
the \universe{} expands.  Thus, ignoring curvature effects for the
moment, the \asd{} is the proper distance to the object at the time the
light was emitted:  The proper distance \Dp{} (sometimes written
$D_{\mathrm{p}}$ or $D_{\mathrm{P}}$) is the distance which one could,
in a \gedankenexperiment{}, measure with a rigid ruler instantaneously
(such that the distance does not change during the measurement due to
the expansion of the \universe{}).  As such, it changes with time due to
the expansion of the \universe{}.  Often, the co-moving distance is
defined as the proper distance at the present time.  Thus, the proper
distance at a different time is simply the current proper distance
divided by ($1+z$), the time being that when the light of an object with
redshift $z$ was emitted. This agrees with the definition used by many
authors, such as 
\citet{MBerry86a},
who defines it as `the distance measured with a standard rod or tape, in
a reference frame where the events occur simultaneously'.  Beware that
sometimes the same distance is denoted by different symbols,
\eg{}~$d_{\mathrm{prop}}$ by 
\citet{SWeinberg72a},
$L$ by
\citet{EHarrison93a},
$d$ by
\citet{ASandage95a},
$D$ by
\citet{TDavisCLineweaver04a},
$d_{\mathrm{p}}$ by
\citet{WHeacox15a},
$d_{p}$ by
\citet{BRyden17a},
and sometimes also by different names, though it is clear from the
discussion that the same distance as that called the proper distance by
\citet{SWeinberg72a} 
is being discussed, \eg{}~`distance between two fundamental particles
at time $t$' ($D_{1}$) by 
\citet{HBondi61a},
`tape-measure distance' ($L$) by
\citet{EHarrison2000a},
`instantaneous physical distance' by
\citet{SCarroll19a}; 
the term `line-of-sight comoving distance' is also sometimes used, as
opposed to the `transverse comoving distance', which is very confusingly
called the `angular size distance' by 
\citet{PPeebles93a}, 
who uses the term `angular diameter distance' for what is called the
\asd{} by almost everyone else\tdash{}indeed, the two terms are usually
considered to be equivalent; the transverse comoving distance is the 
same as the proper-motion distance; see KHS.

\begin{figure}[t]
\vspace{-2cm}
\vspace{1.25em}
\begin{center}
{}\hspace{-1cm}
\includegraphics{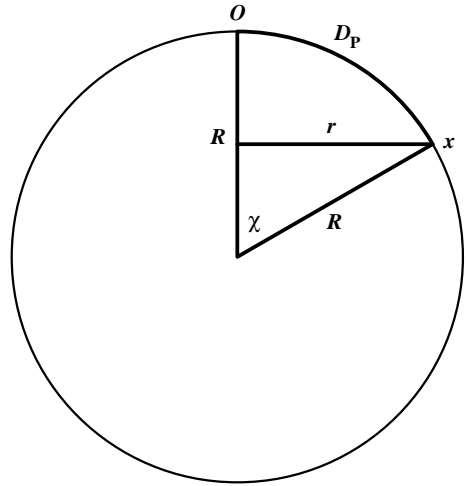}
\end{center}
\vspace{-1cm}
\caption{Although the corresponding definitions are valid for models
with $k$ of $0$ and $-1$ as well, easiest to visualize are distance
definitions for the case $k=+1$.  The \universe{} can be thought of as a
curved three-dimensional space, corresponding to the circle.  Two
dimensions are hence suppressed, so that the two dimensions in the plane
of the figure can show the \universe{} and its spatial curvature.  $R$
is the scale factor of the \universe{}, as usual chosen to correspond to
the radius of curvature.  The observer is located at the top of the
circle at $O$ and observes an object located at $x$.  \Dp{}, the length
of the arc, is the proper distance to that object.  For $\eta=1$, the
angular-size and luminosity distances (as well as other distances not
discussed here such as the proper-motion distance and parallax distance)
depend on $r=R\sin(\chi)$ in a relatively simple manner (see KHS).  Note
that $\chi$ is constant in time; one can use it or $\sigma=r/R$, which
is also constant in time, as the basis for a so-called co-moving
distance.} 
\label{distancedefinitions}
\end{figure}

At small redshifts, as the redshift increases, the object was farther
away (in proper distance) when the light was emitted, thus the \asd{}
increases with redshift.  However, at large redshifts, light was emitted
when the proper distance was small, long ago, but, due to the more rapid
expansion of the \universe{} in the past, is reaching the observer just
now.  Thus, at large redshifts, the \asd{}, being the proper distance
when the light was emitted, is small.  This explains the `remarkable'
maximum.  Another way of thinking of this is that the \asd{} approaches
zero as $z$ approaches 0, but also as $z$ approaches $\infty$, because
the scale factor $R$ (see \fig{distancedefinitions}) approaches 0 in
such cases; in other words, the maximum in the \asd{} depends on a
finite particle horizon.  Of course, not all cosmological models have a
finite particle horizon and those that don't also have no maximum in the
\asd{}.  This applies only to the standard distance, \ie{}~assuming
complete homogeneity.  For the ZKDR distance, it is of course possible
that there is no maximum in the \asd{} even though the \universe{} has a
particle horizon. 

The above explanation is exact in a spatially flat \universe{}, thus
contradicting the claim that the maximum is somehow caused by the
curvature of space.  With spatial curvature, the \asd{} corresponds not
to the proper distance when the light was emitted, but rather to the
coordinate distance $r$, defined as the product of the scale factor $R$
and $\sin(\chi)$, $\chi$, or $\sinh(\chi)$ for $k$ equal to $+1$, $0$,
or $-1$, \ie{}~positive, zero, or negative spatial curvature,
respectively; $\chi=\Dp{}/R$ (see \fig{distancedefinitions}.) 
This is analogous to the correction applied due to the curvature of the
surface of the Earth when calculating the length along a parallel of
latitude from the difference in longitude betweem the ends; the length
(in the limit of small $\theta$) is not $\Dp{}\theta$ but rather
$R{}\sin(\chi)\theta$, where $\chi$ is $\Dp{}/R$, $\Dp{}$ being the
distance measured along the surface of the Earth (`as the crow flies') 
and $R$ is the radius of the Earth (assumed to be perfectly spherical).
(Note that $\chi=\pi/2-\phi$, where $\phi$ is the geographic latitude,
if we think of the observer as being at the north pole.)  Thus, this
distance at first increases with increasing $\Dp{}$, though more slowly
than in the flat case, reaches a maximum at the equator, then decreases
to zero at the opposite pole.%

\Fig{distancedefinitions} illustrates various distances.  One can see
that for small $\chi$, \Dp{} and $r$ are approximately the same (exactly
so in the limit $D_{\mathrm{P}}=r=0)$.  When $\chi$ reaches 90
\degrees{}, $r$ (and hence the \asd{}) reaches its maximum.  For larger
$\chi$, the \asd{} decreases, reaching 0 for $\chi=180\degree{}$.  It
then increases again, reaching its maximum again at $\chi=270\degree{}$,
then decreases again, reaching 0 at $\chi=360\degree{}$.  The maximum
value of $\chi$ depends on the cosmological model.  Light travels along
the circle from $x$ to $O$.  In an expanding \universe{}, $R$ was
smaller when the light was emitted, hence, the distance defined \via{}
light-travel time is smaller than \Dp{}, while they coincide in a static
\universe{}.  The ratio $R_{0}/R_{\mathrm{e}}$, the scale factor now
compared to the scale factor when the light was emitted, is equal to
$1+z$. Distances related to $r$ depend on $\eta$, while \Dp{} and the
distance defined \via{} light-travel time do not (the latter at least to
a very good approximation.)  (More precisely, the \asd{} and other
distances can be calculated relatively easily from $r$ for $\eta=1$. For
$\eta\neq1$, $r$ still exists, but the relation between $r$ and the
angle defining the distance is changed, so the distance can no longer be
simply calculated from $r$.)  For $\eta=1$, since $\Dasd{}=r/(1+z)$, it
is clear that for $z=\infty$ the \asd{} must be zero.  

That is another mechanism for the presence of a maximum in the \asd{}.
Consider first a static \universe{} with positive spatial curvature (the
Einstein model) and an observer at the `pole'.  For increasing proper
distance, the angular size of a standard rod first decreases (\ie{}~the
\asd{} increases) up to a minimum at the `equator', then increases
again, becoming infinite at the opposite `pole'.  (This can continue
indefinitely, with the angular size decreasing again as the proper
distance further increases until the `equator' is reached (but at the
`opposite side'), then increasing again until the object returns back to
the observer, then decreasing again during the second loop around the
\universe{}, and so on.)  Of course, in a static model there is no
redshift, but there are quasi-static models where the \universe{}
expands very slowly.  Large differences in proper distance correspond to
small differences in redshift, and hence small differences in the scale
factor at the time the light was emitted.  If light is received from an
object near the opposite `pole', it will obviously have a much smaller
\asd{} than one near the `equator', even though the scale factor was
only slightly smaller when the light was emitted in the former case
(thus it will have a slightly larger redshift).  (Our \Universe{} never
went through such a quasi-static phase, so the first effect is more
important in practice.) 

As noted above, the claim that the maximum is due to the curvature of
space is strange, as it can exist in a flat \universe{}; in particular,
it exists in the first model considered in the paper, the \EdS{} model,
which is spatially flat.  (Perhaps he meant `\spacetime{}' rather than
`space'; it is not a wrong translation, since the original also has the
Russian word for `space' and not `\spacetime{}'.)  The third point is
more interesting: a maximum exists only if the beam is not empty.  Since
Z64 seemed surprised that the maximum exists, while I have shown above
that it is perfectly natural to expect it, perhaps a better formulation
is that the maximum disappears in the empty-beam case.  I return to this
in \sect{DZ65-summary}. 

We are so used to the redshift $z$ as the principle observable quantity
and proxy for distance that the use of $\Delta=1-1/(1+z)$ is rather
confusing.  It does have the interesting property, though, that it
ranges from 0 at the observer to a maximum of 1 for light emitted at the
big bang.  It follows from the simple definition given by \eqn{8} that
$\Delta=(\omega_{0}-\omega_{1})/\omega_{0}$, but note that `$\omega_{1}$
is the frequency of light received by the observer at time $t_{0}$ and
$\omega_{0}$ the frequency of light emitted by the object at time $t$'.
Usually, `0' refers to the time of reception, but in any case various
quantities almost always have the same indices to refer to the same
times.%
\footnote{There is perhaps \emph{some} justification for using the
subscript 0 to correspond to the time of emission when the frequency is
discussed, since times earlier than the present correspond to a
\emph{higher} frequency, though shorter wavelengths and smaller
quantities based on length; this probably creates more confusion than it
avoids, though.%
}
This is not a misprint, though, since other formulae which follow from
this definition can be shown to be equivalent to more-familiar formulae;
\eg{}~\eqn{9} corresponds to \eqn{B24} in the paper by KHS for
$\onull{}=1$, \ie{}~the \asd{} in the (completely homogeneous) \EdS{}
model, denoted by $f$ in Z64; in modern notation, this is
$cH^{-1}2((1+z)^{-\frac{1}{2}}-(1+z)^{-\frac{3}{2}})$.  There is
something of a misprint in \eqn{2} of Z64: the character before the
exponent `2' in the denominator, which looks like a dagger in a slanted
font, should be $t$, and it is not part of the exponent, \eg{}~correct
is 
\begin{displaymath}
\rho = \frac{1}{6\pi\kappa t^{2}} \mathspace{};
\end{displaymath}
perhaps the lower part of the $t$ has not been printed; this is correct
in the Russian orginal 
\citep{YZeldovich64ao}.
(\Eqn{2} in Z64 follows from the standard definition $\Omega = (8\pi
G\rho)/(3H^{2})$ for $\Omega=1$ and using \eqn{3} to express $H$ in
terms of $t$.) 

The bulk of the paper is the appendix (two and a half pages), which
contains all the equations.  The first one-and-one-half pages are
essentially a non-mathematical summary, but also include several
interesting points. 

\Figrs{3}{ and }{4} are never referred to in the text (neither in the 
translation nor in the original).  \Figr{3} seems to show the case in
which the two ends of an object are multiply imaged, while \figr{4}
seems just to show the definition of an angle.  Note that \figr{6} is
incorrect in that it appears that $f_{1}$ has a maximum for $\Delta<1$;
the \asd{} \emph{never} has a maximum for $\eta=0$ (see \sect{DZ65}). 

The intergalactic medium is said to contain neutrinos and gravitons.
Interestingly, gravitons have a rest-mass of 0 and neutrinos were
believed to as well when the paper was written.  Such particles thus
correspond to a different equation of state ($w=1/3$), though in this
case that is irrelevant since the density is assumed to be negligible.
If the density of such matter is not negligible, then matters become
more complicated.  In general, the term $\rho$ above is $\rho+p$, where
$p$ is the pressure.  In the case of ordinary matter (`dust'), $p=0$,
hence $\rho$ is sufficient.  In the case of the cosmological constant,
which can be thought of as a perfect fluid with $\rho=-p$, the two terms
cancel; the only effect of the cosmological constant on the ZKDR
distance is due to its effect on the expansion history of the
\universe{}.  Other equations of state can in principle be taken into
account in the ZKDR \ansatz{} by including the corresponding $\rho+p$
terms, but in such cases the concept of a single parameter $\eta$ would
be inappropriate, since one would not expect the various components to
clump in the same manner. 

Gravitational lensing is mentioned for the case when there is a galaxy
within the cone, confusingly citing Fritz Zwicky (see below in this
section).  In this case, it is noted that no general expression can be
derived, but that (the equivalent of) the \asd{} as a function of $z$ is
given by a weighted mean, though this is not defined, much less derived.
Though worded somewhat confusingly, it is pointed out that a mass
outside the cone acts, to first order, as a pure-shear gravitational
lens, distorting though not changing the area subtended by the object
and (due to the conservation of surface brightness in gravitational
lensing%
\footnote{Since gravitational lensing conserves surface brightness,
magnification (increase in area) implies amplification (increase in
apparent brightness, \ie{}~energy per time from the source received at
the observer).  In this sense, the terms are interchangeable.  However, 
one or the other term can be more appropriate depending on the 
phenomenon discussed, \eg{}~`amplification' when discussing the change
in apparent magnitude of a lensed source and `magnification' when
discussing the size of an extended source.  In the case of the number of
sources in a certain range of apparent magnitude in a given area of sky,
both effects play a role, and whether there is an increase or decrease
depends on the luminosity function.%
}, 
implicitly assumed here) thus also not changing the apparent magnitude. 

His \eqn{11} looks suspicious because the right-hand side of 1200
appears to be too round a number.  To the same precision, the correct
value is 1184.  (A more precise value is 1184.365.  Of course, this much
precision is not needed, but usually all quoted figures are correct.  If
two significant figures are sufficient, then $1.2\times10^{3}$ would
make more sense.)  Units, not explicitly mentioned, are Mpc.  (Note that
the units of $H$ are `km/sec $\cdot$ Mps', normally written `km/s/Mpc'
or `km/(s$\cdot$Mpc)' or `km\,s$^{-1}$\,Mpc$^{-1}$'.) 

As mentioned above, it is noted that `the function $f_{1}$ increases
monotonically right up to the [particle] horizon ($\Delta=1$) where it
reaches the value 2/5'.  However, the plot of this function in \figr{6}
clearly shows a maximum for $\Delta<1$, after which the value decreases
somewhat. 

The reference 
\citet{FZwicky37c}
is wrongly assigned the year 1927.  That reference contains only a very
short and general discussion on `nebulae as gravitational lenses' and
does not address the phenomena mentioned in the text.  It does say that
a more detailed description will be provided in \journal{Helv. Phys.
Act.}, but that is not the paper in that journal mentioned in reference
4
\citep{FZwicky33a},
a paper in German on various aspects of the redshift of extragalactic
nebulae, which doesn't mention gravitational lensing at all; among other
things, Zwicky points out that the dispersion of velocities of galaxies
in the Coma cluster indicates that the density of dark matter must be at
least 400 times that of luminous matter\tdash{}and, of course, this was
written before 
\citet{FZwicky37c}.
\citet{FZwicky37a,FZwicky37b} 
are the (two short) papers which discuss nebulae as gravitational 
lenses, both cited by 
\citet{FZwicky37c}.

\subsection{Discussion}

Z64 presented analytic formulae for the \asd{} for three cosmological
models: $\onull{}=1$ and $\lnull{}=0$ (\EdS{}) for the values $\eta=1$
(standard distance) and $\eta=0$ (the main result of that work), as well
as for $\onull{}=0$ and $\lnull{}=0$ (the general-relativistic
equivalent of the Milne model; since the density is 0 in this case, the
value of $\eta$ doesn't matter). 

Z64 alerted people to the fact that the standard distances, which assume
complete homogeneity, are perhaps not appropriate, and demonstrated that
effects due to a \universe{} with small-scale inhomogeneities can be
appreciable.  He also introduced the idea of calculating the effect as a
negative gravitational-lens effect, based on simplifying assumptions
rather than calculating it for an analytically soluble (but perhaps less
realistic) case.

\section{Dashevskii \& Zel'dovich (1965)}
\label{DZ65}

\citet[hereafter DZ65]{VDashevskiiYZeldovich65at}%
\footnote{This discussion follows the English translation of the Russian
original 
\citep{VDashevskiiYZeldovich65ao}.%
} 
derived an expression for the \asd{} for the case of a completely empty
beam for arbitrary values of \onull{} ($\lnull{} = 0$ is still assumed).
Compared to Z64, it is more general with respect to the large-scale
cosmological model.  They noted that the expression does not have a 
maximum.

\subsection{Summary}
\label{DZ65-summary}

As noted above, Z64 claimed that the maximum in the \asd{} (in the case
of the \EdS{} model studied) `is caused by the curvature of space due to
the matter filling the universe'.  This is somewhat dubious, since the
\EdS{} model is spatially flat.  DZ65 have a perhaps somewhat better
formulation, claiming that the `effect depends on the bending of light
rays by matter present within the light cone' and assert that `it
follows from this that for objects in whose light cone there is by
chance no matter there should be no minimum angular diameter right up to
the [particle] horizon'.  The claims are true, but one can ask whether
their explanation is the best one.  (As will be discussed in
\sect{DS66-summary}, there is \emph{always} a maximum as long as the
beam is not completely empty, though the emptier the beam, the higher
the redshift of the maximum.) 

In addition to the wider range of cosmological models considered, DZ65
derive the expression \via{} a different, though equivalent, route.  No
analytic solutions are presented, but $f$ and $f_{1}$ are plotted as
functions of $\Delta$ for a few values of \onull{}, and for a few more
values of \onull{}, $\Delta_{\mathrm{max}}$ (the value of $\Delta$ at
which the maximum in the \asd{} for $\eta=1$ occurs) and the values of
$f$ at $\Delta_{\mathrm{max}}$ and $f_{1}$ at $\Delta=1$ are tabulated. 
In addition, there is a column for $\onull{}=\infty$, where
$\Delta_{\mathrm{max}}=0.25$,
$f(\Delta_{\mathrm{max}})=0.65/\sqrt{\Omega}$ and
$f_1(\Delta=1)=1.18/\sqrt{\Omega}$.  This is not
mentioned in the text, but is apparently an approximation for
$\onull{}\gg1$.  I have checked this numerically and found that their
approximation answers pretty nearly.  (Of course,
$\Delta_{\mathrm{max}}$ also depends on \onull{}, though less
sensitively than $f(\Delta_{\mathrm{max}}$) and
$f_1(\Delta=1$).) 

Note that an analytic solution, though a rather complicated one, for the
case $\lnull{}=0$ and $\eta=0$ does exist, first derived by 
\citet{CDyerRRoeder72a}; 
\cf{}~KHS, \eqn{B15}.  For $\lnull{}=0$ and $\eta=1$, the formulae
derived by 
\citet{WMattig58a}
apply.%
\footnote{Mattig gave formulae for $\onull{}>0$ and for $\onull{}=0$.
Not only does one need two formulae, but the formula for $\onull{}>0$ is
numerically difficult for $\onull{}\approx0$ 
\citep{Jpeacock99a}.  
Both can be avoided \via{} a more complicated formula which covers both
cases 
\citep{JTerrell77a}.%
}

Several interesting features are pointed out in the text and/or are
obvious from the figure (if \onull{} is not mentioned, then the effect
is independent of the value of \onull{}): 
\begin{itemize}
\item The \asd{} for $\eta=0$ increases monotonically with redshift. 
\item The \asd{} for $\eta=0$ is less than the light-travel\jdash{}time 
      distance $c(t_{0}-t)$ and larger than the \asd{} for $\eta=1$ (at 
      least for $\lnull{}=0$). 
\item The \asd{} for $\eta=0$ has its maximum value at $z=\infty$.
\item For $\eta=0$, $\D D/dz = 0$ at $z=\infty$.
\item The \asd{} for $\eta=1$ has a maximum at $z<\infty$.
\item The value of the maximum of the \asd{} for $\eta=1$ increases with
      decreasing \onull{}. 
\item The redshift of the maximum of the \asd{} for $\eta=1$ increases
      with decreasing \onull{}. 
\item The \asd{} for $\eta=1$ is 0 at $z=\infty$.
\item Both for $\eta=0$ and $\eta=1$, the value of \Dasd{} at any
      redshift increases with decreasing \onull{}. 
\item For given values of \onull{} and $z$, \Dasd{} for $\eta=0$ is
      always larger than \Dasd{} for $\eta=1$.
\end{itemize}

DZ65 end with remarks on the `validity of the method proposed in the
paper', the validity being guaranteed by the fact that they `are adding
small effects in the linear region'.

\subsection{Remarks}

The title is also confusing, since there is no paper with a similar
title but with `I' instead of `II'.  It is clear from the first
sentence, though, that Paper~I is Z64. 

The theme of confusing notation continues.  What Z64 called $r$, DZ65
call $z$.  While $r$ is often used for a length of some sort, this is
less common for $z$.  Of course, the fact that $z$ is normally used for
the redshift adds to the confusion.  What Z64 called $\Theta$, DZ65 call
$\phi$.  DZ65 adopt the usual convention of using the suffix 0 to denote
the present time, in this case the time of observation and the time the
radiation reaches the observer.  Hence, what Z64 called $\omega_{1}$,
DZ65 call $\omega_{0}$, and what Z64 called $\omega_{0}$, DZ65 call
$\omega_{t}$. 

Criticizing 
\citet{JWheeler58a}, 
DZ65 note that the claim that the maximum occurs only in the case of a
spatially closed \universe{} is wrong. 

I have calculated the values in their \tabl{1}, but in two cases find
different values, namely 0.42 (0.421) instead of 0.40 for
$f(\Delta_{\mathrm{max}})$ for $\onull{}=1/10$, and 0.24 (0.237) instead
of 0.23 for $f_{1}(\Delta=1)$ for $\onull{}=10$.  I suspect that the
former is a misprint while the latter could be as well, or possibly due
to roundoff error in a less accurate numerical calculation.

\subsection{Discussion}

DZ65 presented an integral for the \asd{} for cosmological models with
$\lnull{}=0$ but arbitrary \onull{} for $\eta=0$ and compared the
corresponding distances to those with $\eta=1$.  Although no analytic
solution was presented, DZ65 extended to $\eta=0$ the idea of
calculating distances for various values of \onull{} (though still
setting $\lnull{}=0$).  Around the same time, much more extensive
numerical calculations were done by 
\citet{SRefsdalSdL67a}, 
only for $\eta=1$ but for several values of \onull{} and \lnull{}.

\section{Dashevskii \& Slysh (1966)}

\citet[hereafterDS66]{VDashevskiiVSlysh66at}%
\footnote{This discussion follows the English translation of the Russian
original 
\citep{VDashevskiiVSlysh66ao}.%
}
generalized the method of Z64 and DZ65 to the more realistic case that
the beam is not completely empty, but only for the \EdS{} model.

\subsection{Summary}
\label{DS66-summary}

The empty-beam case is criticized as being too unrealistic, as there
will always be some intergalactic matter; this will mean that there will
always be a maximum in the \asd{}.  DS66 derive, in their \eqn{2}, the
second-order differential equation which is the basis for all further
work in this field 
\begin{equation}
\label{dgl}
\ddot{z} - \frac{\dot{a}}{a}\dot{z} + 4\pi G\rho_{g}z = 0 \mathspace{},
\end{equation}
`which determines the linear distance $z(t)$ between rays', with
$\rho_{g}=\alpha\rho$ (the subscript $g$ refers to the \emph{smooth}
component, considered as a `\emph{{\textbf{g}}as} at zero pressure that
fills all space uniformly' [my emphasis], the rest of the `matter being
concentrated in discrete galaxies'); $a$ is the scale factor and $G$ the
gravitational constant.  Compared to Z64 and DZ65, they allow $\alpha$
(in the notation of KHS, $\eta$) to take an arbitrary value
$0\le\alpha\le1$; $\eta$ is thus completely general.  The cosmological
model is implicit in the term $\dot{a}/a$, in principle allowing one to
study any cosmological model in which $\dot{a}/a$ can be calculated, but
DS66 then restrict themselves to the \EdS{} model for the subsequent
discussion, presenting a completely analytic solution for the \asd{} for
this cosmological model, namely the first unnumbered equation in DS66,
which is a generalization of \eqn{10} in Z64. 

DS66 point out that, for arbitrary $0<\eta\le1$, the \asd{} has a
maximum at finite $z$ and the \asd{} goes to 0 for $z=\infty$.  Also,
the smaller the fraction of homogeneously distributed matter, \ie{}~the
smaller $\eta$, the higher the redshift of this maximum.  Without proof,
it is stated that this result also holds in the case of non-zero
pressure.

\subsection{Remarks}

It is not clear why \eqn{3} is the last numbered equation; perhaps
because the following equations are not referred to in the text (but,
like the others, are of course part of the text).  Also confusing is the
expression $0\le\alpha\le1.1\le{}k\le5$, which should be
$0\le\alpha\le1$, $1\le{}k\le5$.  As in Z64,
$\tilde{f}_{1}$%
\footnote{%
It is unclear why D66 use $\tilde{f}$ while Z64 and DZ65 use $f$.%
}, 
\ie{}~the \asd{} for $\eta=0$, is incorrectly shown as having a maximum
at finite $z$ (a mistake also made by DZ65, though barely perceptibly;
in all cases, these are probably due to the figures having been drawn by
hand).  Also, there should be no inflection in the dashed curve.

\subsection{Discussion}

The generalization to an arbitrary value of $\eta$ is obvious; less
obvious is the relatively simple analytic solution for arbitrary $\eta$
for the \EdS{} model.

\section{Other Papers I}
\label{other1}

(Being discussed in an `other papers' section does not imply that the
paper lacks quality or influence; quite the opposite, in fact.  Rather,
these sections discuss papers which are not directly relevant to the
main theme of this review, but nevertheless played some role in it.) 

\citet{JKristianRSachs66a} 
discuss what I like to call `theoretical observational cosmology' for
very general (\ie{}~anisotropic, inhomogeneous) cosmological models, not
necessarily based on general relativity (GR) (of which the \FRW{}
models\pdash{}homogeneous and isotropic models based on GR\pdash{}are
special cases), mainly for inhomogeneities on the scale of $10^{9}$
\lightyears{} or more (with small-scale inhomogeneities considered to be
smoothed out, \ie{}~in some sense the reverse of the assumptions above).
Many results, after `straightforward, though somewhat tedious'
calculations, are given in terms of series expansions.  A key result is
that the relation 
\begin{displaymath} 
d\,A = r^{2}d\Omega
\end{displaymath} 
(in their notation), where `$d\,A$ is the intrinsic cross-sectional area
of a distant object; $r$ is a measured quantity, the ``corrected
luminosity distance,'' defined by \eqn{19}; and $d \Omega$ is the
meaured solid angle subtended by the distant object' is very general and
holds in all cosmological models, whether or not they are based on GR.
At the time, observations were not good enough that one could be sure
that the \Universe{} is actually very well described by \aFRW{} model,
hence the emphasis on generality and discussion of possible observations
which could be used to determine the many more parameters than those
needed to specify \aFRW{} model. 

\citet{BBertotti66a} 
cites Z64 and DZ65 (erroneously making Dashevskii an author of Z64 as
well), but considers not just the increase in the \asd{} as compared
with the standard \FRW{} case, but also the decrease (corresponding to
amplification) due to the gravitational-lens effect, both strong lensing
and weak lensing, \ie{}~`the small, but distance-dependent, brightening
caused by near galaxies' which leads to a `statistical spread in
luminosity', shown to be proportional to $(\Dasd{})^{3}$ for small
distances.  The main result is an expression for apparent luminosity as
a function of redshift, noting that, in the inhomogeneous case, the
first correction is quadratic in redshift and produces a dimming, but
for higher values of $z$ the brightening due to gravitational lensing
becomes more important.  That expression is for arbitrary \onull{}%
\footnote{As was the custom at the time, this was written in terms of
\qnull{}, \ie{}~$\qnull{}=\onull{}/2$ under the assumption $\lnull{}=0$.
The reason that \qnull{}\pdash{}in general,
$\qnull{}=\onull{}/2-\lnull{}$ (or, as was common at the time,
$\qnull{}=\snull{}-\lnull{}$, where $\snull{}=\onull{}/2$)\pdash{}was
used is that \qnull{} is, after \hnull{}, the next-higher term in series
expansions of observational quantities as a function of redshift
\citep[\eg{}][]{FHoyleASandage56a}%
} 
and arbitrary $\eta$ (called $f$), \ie{}~the case considered by DS66%
\footnote{Since 
\citet{BBertotti66a} 
was submitted around the time that DS66 appeared, presumably the former
was derived independently of the latter and \viceversa{}.%
}
but expressed as a series expansion.  It is also noted that, to first
order, the correction to the Euclidean relation to the expression for
the number of sources brighter than a given apparent luminosity does not
depend on $\eta$. 

\citet{JGunn67a} 
also examined statistical fluctuations due to gravitational lensing, but
in position, not apparent magnitude.  This was done in more detail by 
\citet{TFukushigeMJunichiro94a},
who pointed out that `the distance between nearby photons grows
exponentially because the two rays suffer coherent scatterings by the
same scattering object'. 
\citet{JGunn67b} 
extended the discussion to fluctuations in apparent magnitude.  Feynman,
in a colloquium at Caltech, had discussed a scenario similar to that
discussed by Z64, concentrating on the effects on angular diameters,
apparently not realizing that apparent magnitude would also be affected.
For the topic of this review, the most important result is the
realization that, for large-enough redshifts, average luminosities and
angular sizes will be the same as in the strictly homogeneous case,
because not all lines of sight can be underdense, though there will be a
scatter in their values compared to those in a strictly homogeneous
\universe{}. 
\citet{ABabulMLee91a}
discussed Gunn's formalism in more modern notation, adopting some
simplification and deriving some new analytic results.  Although only
the \EdS{} model was considered (with\pdash{}as extreme
positions\pdash{}a spectrum of mass fluctuations derived from CDM and a
white-noise spectrum), their conclusions probably apply more generally,
namely that the dispersion in amplification due to large-scale structure
is negligible, while that on small scales depends strongly on the nature
of the distribution. 

\citet{SRefsdal70a}
also discussed changes in the apparent luminosity and shape of distant
light sources due to intervening inhomogeneities, but using a numerical
ray-tracing approach rather than the more analytic methods of the
works discussed above.  (As would become clear later, this allows the
effect of very concentrated masses, \eg{}~stars, to be taken into
account, as well as general fluctuations due to galaxies and large-scale
structure.  In other words, it can handle strong lensing as well.)
Ray-tracing simulations were done for a static flat \universe{} (with
all the mass in point massees, \eg{}~$\eta=0$), but the results were
generalized to an interesting collection of cosmological models:
Einstein's static \universe{}%
\footnote{Because it is static, \lnull{} and \onull{} are infinite,
since $\hnull{}=0$ ($\lnull{}=\Lambda/(3\hnull^{2})$ and $\onull=8\pi
G\rho/(3\hnull{}^{2}$)).  Both the density $\rho$ and the cosmological
constant $\Lambda$ are positive, $\Lambda=4\pi G\rho$.%
},
two models with $\lnull{}=0$ ($\onull{}=0.3$ and $\onull{}=2$), and a
model with $\onull{}=0.4$ and $\lnull{}=1.7$ (a spatially closed model
which will expand forever with an antipode at $z\approx4$). 

In retrospect, one conclusion was very prescient: 
\begin{quote}
An interesting aspect of the problem is the possibility of using the
effect to obtain information on the mass distribution in the Universe.
Even if the effect is not observable after some systematic efforts to
detect it, one should be able to determine upper limits on the number of
condensed and massive objects in the Universe. 
\end{quote}

\citet{WPressJGunn73a}
pointed out that (at least for $\lnull{}=0$) if \onull{} is due mainly
to compact objects, then the probability is high that a distant source
will be multiply imaged, independently of the mass of the objects (which
does, of course, set the scale of the image separation).  (At the time,
it was not clear that most of \onull{} consists of non-baryonic matter,
and, since arguments against a substantial density of intergalactic gas
had been presented, it seemed natural to look for the missing matter in
compact objects.)  A more detailed analysis shows that the lack of
dependence on the mass is exact, while the image separation has a weak
dependence on \onull{}.%
\footnote{Since the cross section for strong lensing is proportional to
the mass of the lens, the distribution of masses does not matter.  The
weak dependence on \onull{} is due to the effect of \onull{} on the
\asd{}.%
} 
In contrast to the other papers in this section and that in the next
section, the emphasis is on detecting the scattering masses, not the
influence of those masses on observable properties of the sources.
Nevertheless, the ZKDR distance was used, in particular the extreme
empty-beam case, with the lensing effect of individual clumps explicitly
taken into account.

\section{Kantowski (1969)}

\citet[hereafter K69]{RKantowski69a} 
took a somewhat different approach, using Swiss-cheese models 
\citep{AEinsteinESTraus45a,AEinsteinESTraus46a}.  
These are arguably less realistic than the approximation used in the
papers discussed above, since in these models clumps of matter are
surrounded by voids with $\rho=0$.  However, since these models are
exact solutions of the Einstein field equations, the validity of
approximations used to calculate the \asd{} is not an issue (though, of
course, one can question the validity of this approximation to the
distribution of matter).

\subsection{Summary}

`The Swiss-cheese models are constructed by taking a Friedmann model
($p=\Lambda=0$), randomly removing co-moving spheres from the dust, and
placing Schwarzschild masses at the ``center'' of the holes.'  K69 makes
five realistic assumptions in order to facilitate calculations: the
Schwarzschild radii of the clumps are very small compared to their
opaque radii, the size of the Swiss-cheese hole is much larger than the
opaque radius, the change in the scale factor of the \universe{} is
negligible during the time it takes light to cross a hole, there are
enough holes so that the change in the scale factor is negligible during
the time between interactions with two holes, and the mass density of
the opaque clumps is independent of the clump (though not all clumps
have the same mass). 

K69 calculated the bolometric luminosity, which is inversely
proportional to the square of the luminosity distance.  Since the
luminosity distance is larger than the \asd{} by a factor of $(1+z)^2$,
it is easy to compare his results with those discussed above.  At least
under the assumptions mentioned above, the relation between scale factor
and redshift, $1+z=R_{0}/R$ ($R$ is the scale factor of the \universe{}
and, as usual, the subscript $0$ denotes the currecnt value), holds to a
high degree of accuracy.  (The previous discussions mentioned above
essentially assume, though correctly, that this is the case.) 

K69 used the optical scalar equations 
\citep{RSachs61a}
as the starting point for his calculations, as did 
\citet{JGunn67a}.
These describe the expansion, shear, and twist of the cross section of a
beam of light due to the gravitational effect of matter on the beam, and
are a special case of the Raychaudhuri equation 
\citep{ARaychaudhuri55a}.

Historically, when observational cosmology was done with objects at low
redshift, cosmology was `a search for two numbers' 
\citep{ASandage70a},
$H_0:=\dot R/R$ (giving the scale) and the deceleration parameter
$q_0=-\ddot R R/\dot R^{2}\equiv-\ddot R/(R H^{2})=\onull{}/2-\lnull{}$
(describing the first higher-order effects).  K69 points out that in the
case that most matter is in clumps (\ie{}~$\eta\approx{}0$), a real
value of $\qnull{}=2.2$ would, were one to wrongly assume the standard
distance, appear as $\qnull{}=1.5$.  This foreshadows later work, for
example as discussed in \sect{mzr}, stressing the importance of taking
inhomogeneities into account in classical observational cosmology, at
least as long as a significant fraction of matter is in clumps and the
\Universe{} is similar to the approximations used to calculate distances
in such a case.

\subsection{Remarks}

Both the approach of K69 and that discussed in the previous sections
have clumps of matter embedded in a smooth distribution of matter. 
However, because the Swiss-cheese approach of K69 has the clumps
surrounded by voids, the mass of the clumps being equal to the mass
removed from the voids, the density of the smooth component is the same
as the overall density $\rho$, whereas in the previous approach the
density of the smooth component is $\eta\rho$, while that of the clumps
is $(1-\eta)\rho$.  Thus, light propagating outside the holes will
propagate exactly as in the completely homogeneous case; the defocussing
occurs only because the beam crosses some holes.

\subsection{Discussion}

K69 has been very influential, having at the time of writing more than
180 citations, almost as many as the first three papers discussed above
together, although these papers are easier to understand and good enough
for most purposes; the fact that the latter have fewer citations is
perhaps due to their having been published in (a translation of) a
Russian journal.

\section{Dyer \& Roeder (1972)}

\citet[hereafter DR72]{CDyerRRoeder72a} 
discussed the completely empty-beam case; despite starting out with an
expression for arbitrary \onull{} and \lnull{} (using the standard
notation at the time with $\snull{}=\onull{}/2$ and
$\qnull{}=\snull{}-\lnull{}$), results were presented for
$\snull{}=\qnull{}$, \ie{}~$\Lambda=0$ (and hence $\lnull{}=0$).

\subsection{Summary}

For an integral expression for the \asd{}, analytic solutions are
presented for the three cases $\onull{}<1$, $\onull{}=1$, and
$\onull{}>1$; only the much simpler solutions for $\onull{}=1$ (Z64) and
$\onull{}=0$ 
\citep{WMattig58a}
(see also Z64) were previously known.  As was also pointed out by DZ65,
there is no maximum in the \asd{} for $\eta=0$.  The famous result of 
\citet{IEtherington33a}, 
\begin{equation}
\label{Etherington}
\Dl = (1+z)^2\Dasd{} \mathspace{},
\end{equation}
is invoked to note that an empty beam leads to a lower apparent
luminosity which, as discussed by 
\citet{RKantowski69a},
leads one to underestimate \qnull{} if a completely homogeneous
\universe{} is assumed; their example has a real value of
$\qnull{}=1.82$ which, if calculated assuming a completely homogeneous
\universe{}, results in the value $\qnull{}=1.40$. 
\citet{RKantowski69a} 
had a real value of $\qnull{}=2.2$ being interpreted as $\qnull{}=1.5$.
The exact numbers are not important; the point is that, to first order,
the ZKDR distance is larger than in the standard case, which is also the
case for a lower value of \qnull{}.  But this is only to first order;
with higher-redshift data, the two effects are not degenerate.  It is
also shown that, while the difference between the ZKDR distance and the
standard distance is non-negligible, there is little difference between
the ZKDR distance and that obtained by numerical integration in a
corresponding Swiss-cheese model (which, as mentioned above, is not an
exactly equivalent model).

\subsection{Remarks}

Compared to the papers discussed above, especially the first three,
there is much less emphasis on physical models and more on mathematical
results.  Also, comparisons are done between a relatively simple formula
and a more involved numerical integration based on a more complicated
mass distribution.

\subsection{Discussion}

\citet{CDyerRRoeder72a} 
covered the same ground as DZ65, but more thoroughly, presenting an
analytic solution. 

The distance for an empty or partially filled beam has become known as
the Dyer\ndash{}Roeder distance, although various aspects had been
discussed before.  This is probably due to the fact that the
corresponding papers were published in a major English-language journal,
used standard notation, and were more concerned with results than with
theory.  Dyer and Roeder were certainly responsible for putting the
topic on the agenda of many astronomers.  However, for the reasons
outlined above, I refer to this distance as the ZKDR distance.

\section{Dyer \& Roeder (1973)}

\citet[hereafter DR73]{CDyerRRoeder73a} 
can be seen as a combination of DZ65 and DS66, \ie{}~\onull{} and $\eta$
are both arbitrary (though $\lnull{}=0$ is still assumed).  For the
general case, they derive a hypergeometric equation, and present
explicit solutions for $\eta=$ 0, 2/3, and 1 as well as $\onull{}=0$
(the second one being new).

\subsection{Summary}

As in DR73, general discussion is narrowed down by setting $\lnull{}=0$
before explicit solutions are presented.  Second-order differential
equations for both the \asd{} and the luminosity distance are derived,
though of course once one has a solution one can use the Etherington
reciprocity relation to simply derive one from the other.  Using a
substition, these are converted to hypergeometric equations. 

The special case $\eta=1$ is the solution derived by 
\citet{WMattig58a}
while that for $\eta=0$ is that derived by DR72.  New is a solution for
$\eta=2/3$, which is given for the luminosity distance.  For
$\onull{}=1$, one has the solution derived by DS66, which is given for
the \asd{}.  Differentiation of that equation leads to an expression for
the maximum in the \asd{}, showing that as $\eta$ goes from 1 to 0, the
redshift of this maximum goes from 1.25 to $\infty$.  The point first
made by Z64, that the maximum is due to matter in the beam, is
emphasized.  (Note, however, that an arbitrarily small $\eta$ will lead
to a maximum, though at arbitrarily large $z$.)  They suggest comparing
observations with calculations for each of the three values of $\eta$
for which there is an analytic solution, given the lack of knowledge
about intergalactic matter.  Finally, as in DR72, they note that
calculations for Swiss-cheese models (interestingly, including
$\lnull{}\neq 0$) confirm that this is a good approximation, \ie{}~`the
mass deficiency in the beam is in general much more important than the
gravitational-lens effect for reasonable deflectors', at least for
`redshifts in the range of interest'.

\subsection{Remarks}

There is a huge literature on hypergeometric functions, and many well
known functions, including many used in phyics, are special cases, but
in general it is not possible to reduce hypergeometric functions to
(combinations of) standard functions which are easily and efficiently
calculated, either analyically or numerically.  As such, the fact that
the distance equations are hypergeometric equations is interesting, but
(except for the analyically soluble special cases) of little practical
use.

\subsection{Discussion}

DR73 can be seen as a combination of DZ65 and DS66, \ie{}~\onull{} and
$\eta$ are both arbitrary (though $\lnull{}=0$ was still assumed). 

Starting with the \EdS{} model, Z64 had investigated $\eta=0$,
presenting an analytic solution (as well as one for $\onull=0$, in which
the value of $\eta$ irrelevant since there is no matter).  DZ65 had
expanded this to arbitrary \onull{}, though no analytic solution was
presented.  DS66 had returned to the \EdS{} model, but allowed $\eta$ to
be arbitary.  DR72 had covered the same ground as DZ65, but presented an
analytic solution.  Finally, DR73 addressed the most general case so
far, with both \onull{} and $\eta$ as free parameters, and presented
analyic results (most already known) for special cases.

\section{Dyer \& Roeder (1974)}

\citet[hereafter DR74]{CDyerRRoeder74a}
extended Swiss-cheese models to include cases where $\lnull{}\neq 0$ and
showed that the distances so computed correspond well to those based on
previous work (DR72, DR73).

\subsection{Summary}

After a short review of previous work on the topic, the method of
\citet{RKantowski69a}
is extended to $\lnull{}\neq 0$.  Essentially, $\lnull{}\neq 0$ affects
the expansion history of the \universe{} but nothing else; in
particular, $R_{0}/R=1+z$ still holds.  A second-order differential
equation for (a quantity simply related to) the \asd{} is presented, but
no solution is given.  It is noted that a `series solution about $z=0$
can be obtained', but the emphasis is on calculating the correction
factor relative to the homogeneous model of the same mean density; there
is a series expansion for this, but it breaks down by the time the
redshift has become high enough for the effect to be interesting, so
results have to be calculated numerically.  With regard to distortion of
the beam, they show that a beam retains its elliptical cross section,
though orientation and ellipticity can change. 

In Swiss-cheese models, the structure of the clumps must be taken into
account, but for realistic assumptions (assuming that the clumps model
galaxies), `the calculations indicate that\dots the
distance\jdash{}redshift relations do not differ signifancly from the
``zero-shear'' relations discussed in [DR72 and DR73].  Similarly, the
distortion effect has been found to be negligible in the range of
redshifts observable at present, being at most a few percent.'  Although
the Swiss-cheese models are perhaps unrealistic in that real galaxies
are not usually surrounded by a region of lower than average density,
they do show the potentially real effect that there is a dispersion in
the distance calculated from redshift which increases with redshift. 
(In \sect{S-c} it is discussed how important this is for our
\Universe{}.)  Another important result is that the dependence of the
distance\jdash{}redshift relation on \onull{} is increased for
$\eta\approx 0$, thus reducing the precision obtainable in practice.
Previous conclusions mentioned above that decreasing $\eta$ means that
observations interpreted assuming that $\eta=1$ will underestimate
\qnull{} are repeated.

\subsection{Remarks}

Calculations involving the Swiss-cheese models are inherently
statistical in nature and more complicated than those based on
approximations.  The models are even arguably less realistic.  However,
they are important because, being exact solutions to the Einstein
equations, one does not have to worry about approximations.  The fact
that results are very similar to those based on simpler assumptions is
encouraging, and provides justification for using the simpler approach.
It could of course be the case that this approach is too simple for the
real \Universe{}, but in that case a Swiss-cheese model would also 
probably be too unrealistic.

\subsection{Discussion}

DR74 is interesting because it presents for the first time
distance\jdash{}redshift relations in a \universe{} with arbitrary
\onull{}, \lnull{}, and $\eta$.  However, not only because the
calculations are based on Swiss-cheese models, no closed formulae are
given.

\section{Other Papers II}

\citet{RRoeder75a}
applied the work of DR73 to the data of 
\citet{ASandageEHardy73a}, 
concluding that the value obtained for \qnull{} depends both on
assumptions about (in)homogeneity and on galaxy evolution and suggesting
$\qnull{}>0.5$ if the conclusion of 
\citet{JGottGST74a}
is assumed, namely that $\eta\approx 0$.

\citet{RRoeder75b}
applied the conclusions of DR73 to a claim by
\citet{AHewishRD-S74a}
that there is a lack of small-diameter sources at the largest redshifts,
whereby they assume the standard \asd{}.  If $\eta<1$, then the \asd{}
is larger than otherwise, and if one wrongly assumes $\eta=1$, then one
will underestimate the true physical size of the source.  Thus, an
inhomogeneous \Universe{} is not a possible explanation of that claim;
rather, it would exacerbate the problem.

\section[Further solutions of the ZKDR distance]{Further solutions
(analytic and numerical) of the ZKDR distance}

\subsection{Kayser, Helbig \& Schramm (1997)}

Increasingly general equations (EQ), analytic solutions (AS), and
numerical calculations (NC) had been presented in the 1960s and 1970s
(AS implies EQ) (all but the last two below discussed above): 
\begin{description}
\item[\citet{YZeldovich64at} (Z64)]            $\eta=0$ for the \EdS{}
                                               model (AS) 
\item[\citet{VDashevskiiYZeldovich65at} (DZ65)]$\eta=0$ for arbitrary
                                               \onull{}, $\lnull{}=0$
                                               (NC) 
\item[\citet{VDashevskiiVSlysh66at} (DS66)]    $0\le\eta\le1$ for the
                                               \EdS{} model (AS) 
\item[\citet{RKantowski69a}(K69)]              Swiss-cheese models:
                                               $0\le\eta\le1$, arbitrary
                                               \onull{}, $\lnull{}=0$
                                               (EQ, NC) 
\item[\citet{CDyerRRoeder72a} (DR72)]          $\eta=0$ for arbitrary
                                               \onull{}, $\lnull{}=0$
                                               (AS) 
\item[\citet{CDyerRRoeder73a} (DR73)]          $0\le\eta\le1$ for
                                               arbitrary \onull{},
                                               $\lnull{}=0$
                                               (hypergeometric EQ); AS
                                               for $\eta=(0,2/3,1)$ 
\item[\citet{CDyerRRoeder74a} (DR74)]          Swiss-cheese models:
                                               $0\le\eta\le1$, arbitrary
                                               \onull{} and \lnull{}
                                               (EQ, NC) 
\item[\citet{CDyerRRoeder76a}]                 $\eta>1$ (heuristic, not
                                               exact, approach) 
\item[\citet{RKantowskiVB95a}]                 $0\le\eta\le1$ for
                                               arbitrary \onull{},
                                               $\lnull{}=0$
                                               (hypergeometric EQ); AS
                                               for arbitrary values of
                                               $\eta$ 
\end{description}
The only expression available for $\lnull{}\neq0$ was a complicated
differential equation derived by 
\citet{CDyerRRoeder74a}, 
but for Swiss-cheese models.  No closed solution was presented.  Of
course, it can be integrated numerically.  However, it is rather
cumbersome, and the terms do not have an obvious physical interpretation
like those in the differential equations of Z64 and DS66.  While it was
appreciated that Swiss-cheese models are in some sense equivalent to the
ZKDR distance derived \via{} the Zel'dovich method, this was not shown
strictly until much later 
\citep{PFleury14a}.  Thus, between the work of
\citet{CDyerRRoeder76a}
and 
\citet{RKantowskiVB95a},
work on the ZKDR distance concentrated mostly on understanding the
approximation, applications (both in more-traditional cosmology and in
gravitational lensing), and, to some extent, more-realistic models (this
field would come into its own only later, when computer power allowed
more-complicated scenarios to be investigated).  However, the
development of the basic ZKDR distance picked up again later. 
\citet{RKayser85a}
derived a differential equation for the \asd{} in the style of Z64,
DS66, and DR73, but for $0\le\eta\le1$ and arbitrary values of \lnull{}
and \onull{}, which he integrated numerically \via{} standard but basic
means. 
\citet{RKayserHS97Ra}
saw a need for an efficient numerical implementation of that equation,
which is the most general equation for the ZKDR distance under the
standard assumptions that the \universe{} is a (just slightly) perturbed
\FRW{} model (\ie{}~no pressure, no dark energy more complicated than
the cosmological constant, no back reaction, only Ricci (de)focussing;
even today, there is no evidence that the first three are not excellent
approximations, and the fourth is as well in many cases).  Also, no
efficient general implementation existed for the standard ($\eta=1$)
distance.%
\footnote{Despite having been known for decades that the standard
distance can be calculated \via{} elliptic integrals, this was almost
never done, results being presented as a `qualitative integration' 
\citep[\eg{}][]{HBondi61a}
or calculated numerically
\citep[\eg{}][]{SRefsdalSdL67a}.%
}
Thus, a description of the differential equation derived by 
\citet{RKayser85a}
and the efficient numerical implementation\pdash{}using the
Bulirsch\ndash{}Stoer method in \code{Fortran} \citep[see][for technical
details]{Phelbig96Ra}\pdash{} evolved to include a general description
of various types of cosmological distances and a compendium of analytic
solutions, probably the first time all this information had been
presented in a uniform notation.  Despite being a numerical (though very
efficient) implementation, it is only a factor of $\approx3$ slower than
elliptic-integral solutions for $\eta=0$ or $\eta=2/3$ (Rollin Thomas,
personal communication); for $\eta=1$, the factor is $\approx20$ 
\citep{RKantowskiKT00a}.  
Of course, a comparison can be done only for those cases where
elliptic-integral solutions exist, but the numerical-integration time
for the differential equation, valid for all values of \lnull{},
\onull{}, and $\eta$, is essentially the same whether or not an
elliptic-integral or analytic solution exists.  (Analytic solutions are
of course faster than elliptic-integral solutions, which can be
described as semi-numerical or semi-analytic; in general, the
elliptic-integral solutions do not work if there is an analytic solution
(an exception being the expression for light-travel time in a flat 
\universe{}).)

\subsection{Kantowski and collaborators}

Kantowski, with collaborators, had returned to the topic of distance
calculation in locally inhomogeneous cosmological models 
\citep{RKantowskiVB95a},
coincidentally around the same time that I was writing the code for KHS.
Although partially motivated by the \mzr{} for \tIas{}, further progress
was made regarding the theory. 
\citet{RKantowski98b}
used the Swiss-cheese formalism to derive an analytic expression for the
ZKDR distance using Heun functions, valid for arbitrary \lnull{},
\onull{}, and $\eta$. 
\citet{RKantowskiKT00a}
gave analytic expresssions using elliptic integrals for arbitrary
\lnull{}, \onull{}, and $\eta=(0,2/3,1)$, corresponding to
$\nu=(2,1,0)$; see \eq{nu} in \sect{mzr}.  For the flat-\universe{}
case, there are simpler expression involving associated Legendre and
hypergeometric functions; these were given by 
\citet{RKantowskiRThomas01a}.
\citet{RKantowski03a}
pointed out that the general case can be expressed via the Lam\'{e}
equation, which can be solved \via{} Weierstrass elliptic integrals for
$\nu=(2,1,0)$.  While not directly related to the ZKDR distance, but
related mathematically, note that 
\citet{RThomasRKantowski00a}
also expressed the age\jdash{}redshift relation (related to lookback
time and light-travel\jdash{}time distance) via incomplete Legendre
elliptic integrals, but only for $\lnull{}>0$.%
\footnote{The history of the use of elliptic integrals to calculate
cosmological quantities is interesting in itself, but is beyond the
bounds of this review.%
}
To summarize:
\begin{description}
\item[\citet{RKantowski98b}]        arbitrary \lnull{}, \onull{}, and
                                    $\eta$ using Heun functions
\item[\citet{RKantowskiKT00a}]      arbitrary \lnull{}, \onull{}, and 
                                    $\eta=(0,2/3,1)$ (corresponding to
                                    $\nu=(2,1,0)$) using elliptic
                                    integrals 
\item[\citet{RKantowskiRThomas01a}] flat but otherwise arbitrary;
                                    associated Legendre and 
                                    hypergeometric functions
\item[\citet{RKantowski03a}]        Lam\'{e} equation for the general 
                                    case, Weierstrass elliptic integrals 
                                    for $\eta=(0,2/3,1)$ (corresponding
                                    to $\nu=(2,1,0)$) 

\item[\citet{RThomasRKantowski00a}] age\jdash{}redshift \via{} incomplete 
                                    Legendre elliptic integrals
\end{description}

\section{Testing the approximation}
\label{testing}

Unlike the Swiss-cheese model, the ZKDR distance is an approximation
based on various assumptions.  While it is reasonably clear that it must
be correct in the appropriate limit (\ie{}~the light propagates
\emph{very} far from all clumps, the fraction of mass in clumps is
negligible so that it is clear that \aFRW{} model is a good
approximation, \etc{}), it is not immediately clear how good the
approximation is in a more realistic scenario.  One way to test this is
to compare the ZKDR distance to an explicit numerical calculation,
namely following photon trajectories through a mass distribution
produced by a cosmological simulation.  Some of this work will be
mentioned below in \sect{more-detailed}. 
\citet{KWatanabeKTomita90a},
building on work by
\citet{TFutamaseMSasaki89a},
solved directly the equations of null geodesics and explicitly
calculated the shear.  Only the \EdS{} model was considered, and the
explicit calculations were compared to the ZKDR distance for $\eta=1$
and $\eta=0$.  The former is the better fit for the \emph{average}
distance, but it was assumed that mass is transparent, so this result
essentially follows from flux conservation 
\citep{SWeinberg76a}. 
\citet{MKasaiFT90a}
carried out a similar study, noting that, as expected, the
distance\jdash{}redshift relation depends on angular scale, with the
standard ($\eta=1$) distance appropriate for large angles and the ZKDR
distance (in the limiting case, $\eta=0$) for small angles, a
conclusion also arrived at by 
\citet{ELinder98a}.
His numerical result was demonstrated analytically by 
\citet{KWatanabeKTomita91a}.
Similar results were found by 
\citet{JGiblinMS16b},
who used a much more realistic model of the mass distribution, based on
state-of-the-art simulations (`the first numerical cosmological study
that is fully relativistic, non-linear and without symmetry') 
\citep{JGiblinMS16a,JMertensGS16a}.  
They stressed the scatter in the distance for a given redshift, which
generally increases with redshift and is also dependent on the line of
sight. 
\citet{TNakamura97a}
numerically investigated the effect of shear on the \asd{} in a linearly
perturbed \FRW{} model and found it to be negligible, thus justifying
the ZKDR distance.  (For the \EdS{} model, an analytic result was
presented.) 
\citet{TOkamuraTFutamase09a},
while not setting out to test the ZKDR distance, found that a
\universe{} with the halo-mass function of 
\citet{RShethGTorman99a}
is, remarkably, well approximated by the ZKDR distance with the $\eta$
parameter calculated from their model. 
\citet{VBustiHC13a}
compared the ZKDR distance to other approximations: the weak-lensing
approximation with uncompensated density along the line of sight, the
flux-averaging approximation, and a modified ZKDR distance which allows
for a different expansion rate along the line of sight.  This work is
interesting for its analysis of the underlying issues (essentially
assumptions about the mass distribution and how this affects light
propagation, different approximations corresponding to different
assumptions) and its combination of detailed theory and application to
real data\tdash{}the Union2.1 sample, also used by 
\citet{PHelbig15Ra}
and
\citet{XYangYZ13a}.

\section{Weinberg (1976)}

\citet{SWeinberg76a}
pointed out that in a locally%
\footnote{In the context of the ZKDR distance, `locally' means `on small
scales', not some local inhomogeneity near us in an otherwise (more)
homogeneous \universe{}.%
}
inhomogeneous \universe{} in which gravitational deflection by
individual clumps is taken into account, the conventional distance
formulae remain valid \emph{on average} as long as the clumps are
sufficiently small, while for galactic-size clumps, this depends on the
selection procedure and redshift of the source.

\subsection{Summary}

For a locally inhomogeneous \universe{}, the \emph{average} apparent
luminosity (for the case $\lnull{}=0$, but this is true in general) is
given by the conventional formula, \eg{}~that due to 
\citet{WMattig58a},
rather than the empty-beam formula, \eg{}~that investigated by 
\citet{CDyerRRoeder72a}.
The reason is clear: the empty-beam formula `leaves out the
gravitational deflections caused by occasional close encounters with
clumps near the line of sight'.  Moreover, `[t]hese gravitational
deflections produce a shear which on the average has the same effect in
the optical scalar equation as would be produced in a homogeneous
universe by the Ricci tensor term'. 

The special case of $\qnull{}\ll1$ is considered, in which the average
number of clumps close enough to the line of sight to produce an
appreciable deflection is of order \qnull{} for $z\approx 1$ 
\citep{WPressJGunn73a}.  
Even in this case, where multiple deflections can be ignored, the
standard formula is appropriate when considering the average distance.
The decrease in the luminosity distance due to gravitational lensing
cancels the increase due to the empty-beam formula. 

This result is generalized to models with arbitrary \qnull{} and
transparent intergalactic matter \via{} a simple argument: due to flux
conservation, the conventional distance must hold, on average; not all
lines of sight can be underdense, and occasional lines of sight with
strong amplification due to gravitational lensing exactly balance the
larger number of underdense lines of sight.  However, this ignores the
selection effect that there can be no opaque clump between the source
and the observer.  If the clumps are dark stars, the conventional
distance formula is a very good approximation, but only marginally so
for galaxy-size clumps.  The important quantity is the radius of
avoidance, which could lead to the empty-beam distance being more
appropriate at low redshifts and the conventional formula at high
redshifts.%
\footnote{This is a special case of $\eta = \eta(z)$, discussed by KHS.%
}  
Details depend on selection effects: perhaps distant objects are
observed (by accident or by design) on lines of sight which avoid clumps
(and hence absorption); on the other hand, amplification bias might
cause objects which have been gravitationally amplified to be observed
preferentially. 

The empty-beam distance is nevertheless useful since it gives a lower
limit on the apparent luminosity (for a given absolute luminosity) at a
given redshift.  In general, there is a scatter in luminosity distance,
comparable to the difference between the empty-beam and filled-beam
formulae.  Also, it is noted that the standard distance should be used
to calculate the mean inverse-square luminosity distance, not the mean
luminostiy distance itself.  Weinberg speculates that this might be part
of the reason for the difference in apparent luminosity between quasars
at the same redshift.

\subsection{Remarks}

\citet{SWeinberg76a}
is not concerned with developing the theory of the ZKDR distance; in
fact, he doesn't go beyond DZ65.  Rather, the emphasis is on
understanding the validity of the approximation, its domain of
applicability, and its use in a statistical context.

\subsection{Discussion}

This paper has been cited many times, perhaps because Weinberg is well
known, but probably mainly because it is clear and to the point.  Not
until much later were more-detailed analyses presented.

\section{Other Papers III}
\label{otherIII}

\citet{JWArdleRPotash76a}
discussed the effect of the ZKDR distance on the angular sizes of
quasars, noting `that the median angular size in fact decreased with
redshift faster than expected in any Friedmann cosmology.  This implied
that there was a deficiency of sources of large \emph{linear} size at
high redshifts' [emphasis in the original].%
\footnote{As mentioned above,
\citet{AHewishRD-S74a}
arrived at the opposite conclusion.%
}
A cosmological model with $\eta<1$ could at least partially explain
this. 

\citet{RWagoner77a}
discussed determining \qnull{} from the \mzr{} for supernovae, noting in
passing that the Dyer\ndash{}Roeder distance can be used.  While not
dwelling on the question of distance calculation, the paper is one of
the first to advocate determining cosmological parameters from the
\mzr{} for supernovae rather for galaxies, deemed to be worth pursuing
mainly because of the lack of knowledge about galaxy evolution. 

\citet{GEllis78a}
noted that the uncertainty in $\eta$ needs to be considered when
attempting to derive cosmological parameters from observations.  Ellis
would later return to this topic many times.

\section{The end of an era}

The work by 
\citet{SWeinberg76a} 
marks a turning point, for two related reasons.  First, the theory is
now more or less complete; future work would be concerned with
refinements.  Second, the development of theory is now secondary to
applications, at least in terms of numbers of papers.  The three papers
mentioned in \sect{otherIII} are in some sense obvious consequences of
the theory as known when they were written; most future work would be
more limited in scope but also more detailed.  As such, it makes sense
to switch from the mainly chronological discussion presented until now
to a discussion based on topic.  (Nevertheless, some chronology is
retained: topics are presented in the order of their appearance, and the
discussion of each topic is roughly chronological.  The order of the
topics is based not on the average age of the papers, but rather on the
time of publication of the first one.)  Though some build on somewhat
earlier work (some of which has been mentioned above), most of these
topics were investigated after the work of 
\citet{SWeinberg76a}. 

Before doing so, however, the influential work of
\citet{CCanizares82a} 
deserves special mention.  Building on the work of 
\citet{WPressJGunn73a},
who had concentrated on the production of multiple images by compact
objects, he discussed other observational effects.  As such, this work
belongs more in the gravitational-lensing camp than in the
light-propagation camp.  It also appeared at a time which saw a rapid
increase in the number of papers devoted to these two topics. 
Obviously, the discovery of the first gravitational-lens system by 
\citet{DWalshCW79a}
played a role as far as gravitational lensing itself was concerned; but
probably because gravitational lensing forces one to think about the
degree of homogeneity between source and observer, many studies were
done which looked at further applications of the ZKDR distance, and,
somewhat later, refinements to and extensions of the basic theory were
investigated. 

The next 16 sections, discussing various applications of the ZKDR
distance, are chronological with respect to the first paper discussed in
each.  These are followed by a section discussing analytic
approximations; the final section is a summary.

\section{Flux conservation 1}
\label{flux1}

\citet{SWeinberg76a}
pointed out that the standard distance formula, \eg{}~assuming $\eta=1$,
must hold \emph{on average} if lenses are transparent and there are no
selection effects.  This is due to flux conservation. 
\citet{CDyerRRoeder81a}
considered the effect of a finite source size in gravitational lensing,
concluding that, all else being equal, $\eta$ increases with the size of
the source.  (The fact that almost all beams are underdense and hence
the average magnification is less than 1 is offset by the occasional
strong-lensing event.)  The important quantity is not the size of the
source \perse{}, but rather the size of the source relative to the
clumps; as already mentioned by 
\citet{SWeinberg76a},
one could think of $\eta$ increasing with redshift since, due to
structure formation, matter was more uniform at high redshift.  The fact
that the angular size of the beam also increases with redshift (the base
of the cone is at the source; the apex at the observer) is an additional
effect in the same direction.  This was made more explicit by 
\citet{CDyerRRoeder81b},
who showed that, `[i]n the weak-field approximation, the net
amplification resulting from small amplifcations due to many small
spherical deflectors bending light at their perimeters corresponds to
the Ricci amplification where the source and observer are located well
outside the lens'. 
\citet{JEhlersPSchneider86a}
question several assumptions regarding the derivation of the ZKDR
distance.  Subsequent work has shown these doubts to be misplaced;
provided that the \universe{} has a `ZKDR-style' mass distribution, the
ZKDR distance is appropriate.  When calculating the probabilities of a
source being lensed, however, they point out that a random line of sight
is not an average line of sight.  Rather, what is random is the position
of a source on the celestial sphere.  They conclude that lensing
probabilities had thus been underestimated.  This conclusion was arrived
at considering flux conservation for an ensemble of lenses; 
\citet{THamana98a}
showed that it holds for individual beams also (see \Sect{general}). 
The general idea when considering averages is that most lines of sight
are underdense and this is offset by the occasional strong-lensing
event.  In other words, the fact that the average amplification is 1
depends on the existence of an ensemble.  On the other hand, a
transparent lens neither creates nor absorbs photons. 
\citet{YAvniIShulami88a}
showed by an explicit calculation that this also holds for a single,
isolated Schwarzschild gravitational lens; the usual amplification for
small impact parameters is exactly compensated by de-amplification for
large impact parameters. 

Around the same time,
\citet{JPeacock86a}
noted that the solution given by
\citet{CDyerRRoeder73a} 
for arbitrary \onull{} and $\eta$ (but $\lnull{}=0$) is mathematically
valid for $\eta<25/24$, although $\eta>1$ is unphysical, since this
would imply that light propagates along a uniformly overdense tube.
Nevertheless, this can be used as a rough model for gravitational
lensing 
\citep[see also][]{CDyerRRoeder76a}.  
More importantly,
\citet{JPeacock86a}
generalized the result of
\citet{SWeinberg76a}
to arbitrary \onull{}.  (As far as I know, no-one has repeated this
calculation for arbitrary \lnull{}). 
He also agrees that the conclusion of
\citet{JEhlersPSchneider86a}
that a more exact treatment reveals that lensing probabilities had been
underestimated, but points out that their final result is not very
useful since any difference between it and previous estimates becomes
significant only at large optical depth, where the single-lens
approximation breaks down.  (Nevertheless, it still holds that previous
estimates had underestimated the lensing probability.) 

\citet{LFangXWu89a}
pointed out that flux conservation can be used as a constraint when
evaluating various approximations used in calculating the probability of
lensing. 
\citet{JIsaacsonCCanizares89a}
compared the approach of
\citet{WPressJGunn73a}
to that of 
\citet{JEhlersPSchneider86a}
in the \EdS{} model, finding that the former approach can be made to
agree with the latter `by adjusting the average magnification along a
random line of sight so as to conserve flux'. 
\citet{MJaroszynsiBPaczynski96a}
considered flux conservation within the context of microlensing (in
which case $\eta=0$ is appropriate, as long as any smooth mass
distribution is ignored, since the lensing effect is taken into account
explicitly in the microlensing calculation, as opposed to $\eta\approx1$
which would be appropriate if one considered the average effect for a
source size larger than that of the lenses).  They pointed out that in
addition to the redistribution of flux, there is another redistribution
of energy because some observers see an additional redshift, some an
additional blueshift.

\section{Kibble \$ Lieu (2005)}

\citet{TKibbleRLieu05a}
also contributed significantly to the understanding of flux conservation 
in the context of the ZKDR distance; so much so that they deserve their 
own section.  They showed analytically that, under very general
conditions (including arbitrary shapes of clumps and strong lensing),
the average \emph{reciprocal} magnification in a clumpy \universe{} is
the same as that in a homogeneous \universe{}, as long as the clumps are
uncorrelated.  The reciprocal magnification has the advantage that it 
goes to zero rather than infinity on the caustics (regions of---for a 
point source---infinite magnification), and so is more useful in the 
strong-lensing case.  They also discussed various measures of
magnification and the circumstances in which they are appropriate. 

An important distinction is whether one averages over a set of sources
on the unperturbed celestial sphere, or whether one averages over all
lines of sight: `If one part of the sky is more magnified,\dots the
corresponding area of the constant-$z$ surface will be smaller, so fewer
sources are likely to be found there.  In other words, choosing a source
at random will give on average a smaller magnification or larger
angular-size distance.'  This is related to whether it is the mean
magnification or the mean reciprocal magnification that is the same as
in the homogeneous case.  In the weak-lensing case, both are.  In the
strong-lensing case, it is the magnification which averages to 1 over
the celestial sphere, the random-source average\pdash{}the case
implicitly considered by 
\citet{SWeinberg76a}\pdash{}, 
however strong lensing effects are, while it is the \emph{reciprocal}
magnification which averages to 1 over all lines of sight, again however
strong the lensing effects are.  As a corollary, the random-source
average of the \emph{total} magnification of \emph{unresolved} images is
the same as in the homogeneous case, while for \emph{resolved images} it
can be significantly different, essentially because there can be more
than one image of a given source. 

Another distinction is between the \asd{} and the so-called area 
distance (though both distances can be applied to both lengths and 
areas) as introduced by
\citet{GEllisetal98a}.
If strong lensing is involved, \ie{}~multiple images (whether resolved
or not) are present, then the magnification can be defined as negative
for images of odd parity; sometimes, the \asd{} itself is considered to
be negative in such cases.  (This is also the case for an object located
at a coordinate distance $\chi$ between $n\pi$ and $2n\pi$, where $n$ is
an integer, because the rays defining the angle in the definition of the
\asd{} (see \Sect{Z64-remarks}) cross between source and observer.) 
Such areas are counted negatively when calculating the average \asd{};
if the absolute values are used, the corresponding distance is the area
distance, which is thus always larger than the \asd{}.  The area
distance is thus appropriate if one is interested in the total number of
images within a given area of sky or their average magnification; the
\asd{} is appropriate if one is interested in the total number of
distinct sources (say, when multiple images are not resolved) or their
average magnification. 

The work of
\citet{TKibbleRLieu05a}
is also important because it is analytic (though some assumptions are
made, which in practice are always fulfilled to a very good
approximation: the surface of constant $z$ is the same as the surface of
constant affine parameter; shear vanishes when light is propagating far
from all clumps; the clumps are widely separated, slowly moving, and
randomly distributed).  Their work confirms that of 
\citet{SWeinberg76a},
which is based on energy conservation, when averaging over the celestial
sphere (\ie{}~the source is random), and also considers the case of
averaging over lines of sight.

\section{Flux conservation 2}

\citet{YWang00a}
suggested that flux conservation justifies the use of the standard
distance in the analysis of the \mzr{} for \tIas{} and performed such
flux averaging by combining data in redshift bins, pointing out that
this reduces systematic uncertainties from effects such as weak lensing,
while 
\citet{ABarber00a}
claimed that weak-lensing effects are about an order of magnitude larger
than previously found (and hence probably need to be taken into account
more explicitly).  On the other hand, 
\citet{YWang05a}
found only marginal evidence for weak-lensing effects in the \mzr{} for 
\tIas{}.

Even if the \emph{mean} magnification is 1, due to the skewness of the
distribution, the \emph{median} magnification is $<1$. 
\citet{CClarksonetal12a}
pointed out that most narrow-beam lines of sight are significantly
underdense, even for beams as thick as 500\unit{}kpc.  On the other
hand, they also point out that this does not necessarily lead to a
increase in apparent magnitude (\ie{}~dimming) if one drops the
assumption that inhomogeneities can be modelled as perturbations on a
uniformly expanding background, a point also emphasized by 
\citet{KBolejkoPFerreira12a};
see also
\citet{SBagheriDSchwarz14a}.

Although the basic idea of flux conservation is clear (and there are
obvious caveats such as non-transparent matter), exact treatments can be
very complicated and have led to confusion, much of which has been 
cleared up by 
\citet{NKaiserJPeacock2015a}:
\citet{SWeinberg76a}
is essentially right, though one needs to keep in mind the distinction
between magnification and reciprocal magnification as discussed above in
connection with 
\citet{TKibbleRLieu05a}.
Since $\eta\sim\kappa$, where $\kappa$ is the convergence, and
$\mu\sim(1-\kappa)^{-1}$, the relation is linear only in the limit of
vanishing deviations, though approximately linear for the small
deviations considered here.%
\footnote{See 
\citet{YWangTF05a} 
and
\citet{KBolejko11a}
for details on the relationship between the inhomogeneity parameter
$\eta$ and the convergence $\kappa$.%
}
Non-linear functions of the conserved quantity $\mu$ must be handled
with care.  For example, the average \asd{} $\langle\Dasd{}\rangle$, and
hence the average luminosity distance $\langle\Dl{}\rangle$, is biased
even in the case of $\langle\mu\rangle=1$.  This can be considered one
aspect of the averaging problem: we are interested in the average values
of the cosmological parameters determined by observers throughout the
\universe{}, but can at best average observations over several lines of
sight.  See also 
\citet{CBonvinetal15a},
who point out that the ensemble average and the directional average do
not commute; `observing the same thing in many directions over the sky
is not the same thing as taking an ensemble average'; this is a
restatement of the result of 
\citet{TKibbleRLieu05a}.

\section[Galaxy clusters and Swiss-cheese models]{Galaxy clusters and
further work on Swiss-cheese models} 
\label{S-c}

\citet{VRubinFR73a}
noted a non-random distribution of radial velocities on the sky for a
sample of galaxies, later known as the Rubin\ndash{}Ford effect, and
discussed various possible explanations, though none involving
gravitational lensing in any form. 
\citet{HKarojiLNottale76a}
confirmed the effect with two samples of galaxies chosen from the
literature, discussed a number of possible causes, and tentatively
concluded that `light emitted by distant galaxies are [\sic{}]
redshifted when passing through clusters of galaxies or distant sources
are more luminous when seen through intermidiate clusters of galaxies
which could act as gravitational lenses'.  Similar work was done by 
\citet{LNottaleJVigier77a}.
\citet{CDyerRRoeder76a}
tried to explain the Karoji\ndash{}Nottale effect via $\eta>1$.  On the
one hand this is straightforward: $\eta<1$ implies that there is less
matter in the beam than for a random line of sight, so $\eta>1$ would
imply that there is more.  On the other hand, this situation violates
the assumptions under which the ZKDR distance is calculated, so the
applicability is somewhat questionable.  In any case, the conclusion was
that accounting for the effect of gravitational lensing by clusters of
galaxies in this manner cannot explain the Karoji\ndash{}Nottale effect.
Swiss-cheese models were also used to estimate the effect of
inhomogeneities on the CMB 
\citep[\eg{}][]{CDyer76a,LNottale84a},
but this strays too far from the main topic of this article.

\citet{LNottale82a},
in the spirit of
\citet{RKantowski69a},
developed a more complicated but exact-solution model; the question is
then how realistic it is physically, rather than whether the
approximations are valid.  While the Swiss-cheese model of 
\citet{RKantowski69a}
had holes consisting of completely empty voids with the mass removed
from the void concentrated at the centre, and the corresponding
Schwarzschild volume considered opaque, 
\citet{LNottale82a}
had a more realistic model where the mass removed from the hole forms a
Friedmann model of higher density than that surrounding the hole;
importantly, the matter at the centre of the hole is transparent.
Between the two Friedmann solutions is a Schwarzschild solution.  The
main conclusion here is that there is a change in the observed redshift
of objects seen through such a cluster. 
\citet{LNottale82b} 
examined the perturbation of the magnitude\jdash{}redshift relation in
that model, deriving an expression for the change in magnitude dependent
on the cosmological model (\hnull{},\qnull{}), $\eta$, the cluster
radius, the cluster redshift, and the source redshift; typical values of
those parameters result in `some tenths of magnitude'. 
\citet{LNottale83a}
studied this model with respect to `the effects intrinsic to a cluster,
\ie{}~the purely gravitational perturbations on redshift and magnitude
(or equivalently diameter) for sources situated \emph{in} a cluster,
with respect to exterior sources' [emphasis in the original]. 
\citet{LNottaleFHammer84a}
investigated this in more detail, examining the amplification of light
from distant sources by a transparent lens \via{} an exact solution of
the optical scalar equations 
\citep{RSachs61a}.
\citet{LNottaleBChauvineau86a}
used this formalism to calculate the global Ricci amplification by
multiple gravitational lenses, noting that it usually differs
significantly from the product of individual amplifications (an
approximation valid only if all amplifications are small). 

\citet{HSato85a}
continued working with the Swiss-cheese paradigm, finding that the
modification is third order in $Hr_{\mathrm{b}}/c$ for redshift and
first order for apparent luminosity, where $r_{\mathrm{b}}$ is the
radius of a void (Swiss-cheese hole). 
\citet{CDyerLOattes88a}
examined the dispersion of observational quantities such as magnitudes
(related to the luminosity distance) in a Swiss-cheese model,
emphasizing a fundamental limit `due to the ``fuzzy'' structure of the
perceived past null cone' and selection effects due to the skewness of
the distribution of observational quantities (even though the means are
the same as for \FRW{}). 
\citet{NBrouzakisTT08a}
arrived at similar results, noting even `inhomogeneities with sizes of
order 10\unit{}Mpc or larger' cannot lead to `dispersion and bias of
cosmological parameters derived from the supernova data' large enough
`to explain the perceived acceleration without dark energy, even when
the length scale of the inhomogeneities is comparable to the horizon
distance'. 
\citet{TCliftonJZuntz09a}
investigated the effect of large-scale structure on the Hubble diagram
\via{} a Swiss-cheese model. 
\citet{VKostov10a}
examined flux conservation in the sense of averaging over all lines of
sight in Swiss-cheese models, with exact, non-perturbative calculations
including all non-linear effects. 

\citet{PFleuryDU13b}
suggested that the well known `tension' between \satellite{Planck} and
the \mzr{} for \tIas{} 
\citep[see \eg{}~][for \tIas{} data]{AConleyetal11a}
could be relieved if the calculations are done with a Swiss-cheese
model.  This is because the CMB data have a typical angular scale of 5
\arcminw{} while the typical angular size of a supernova is $10^{-7}$
\arcsecw{}.  If the Swiss-cheese model is more appropriate, but a
homogeneous model assumed, then one will underestimate \onull{}.  Note
that at the distances used to determine \hnull{}, the effect of $\eta<1$
is negligible (and would also go in the opposite direction: compared to
$\eta=1$, distances would be larger and hence the derived value of
\hnull{} smaller%
\footnote{%
\citet{IOdderskovKH16a},
by examining the \rdr{} of mock sources in N-body simulations, concluded
that local inhomogeneties cannot explain the tension.  However, they
were looking at the effect on \hnull{} itself.%
}%
).  
Rather, 
\citet{PFleuryDU13b}
pointed out that a lower $\eta$ has, to first order, the same effect as
a lower value of \onull{} (or a higher value of \lnull{}).%
\footnote{To first order in $z$, the luminosity distance dependes on
$z$, to second order on \qnull{} ($\onull{}/2 - \lnull{}$), and to third
order on \onull{} as well as \qnull{}; thus, at relatively low redshift
it is expected that \onull{} is more important than \lnull{} 
\citep[\eg{}][]{JSolheim66a}.%
} 
Thus, incorrectly assuming $\eta=1$ leads to an underestimate of
\onull{}.  If in fact $\eta<1$, then the derived value of \onull{} will
be larger, while the value of \hnull{} changes only slightly.  This
reduces the tension between the values derived by \satellite{Planck} and
the \mzr{} for supernovae, though by changing the value of \onull{}
derived from the \mzr{}.  While the \mzr{} still prefers higher values
of \hnull{}, there is no longer any serious discrepancy with the
\satellite{Planck} results.  This interesting result is a consequence of
the very detailed Swiss-cheese calculations by 
\citet{PFleuryDU13a}.
Alas, as pointed out by
\citet{MBetouleetal14a},
it appears that the low value of \onull{} obtained by
\citet{AConleyetal11a}
was due to a wrong calibration of the MegaCam zero points in the $g$ and
$z$ bands and corrections to the MegaCam $r$ and $i$ filter bandpasses,
thus the analysis by 
\citet{PFleuryDU13a}
is in some sense no longer relevant (though one could turn it around and
see the lack of tension in \onull{} as evidence against such extreme
Swiss-cheese models).  Although interesting because they are exact 
solutions to the Einstein equation, Swiss-cheese models are today
arguably mainly of historical interest.  In particular, the redshift
aspects should not be worrying, since they are merely one aspect of the
integrated Sachs-Wolfe effect, which can be calculated for a CDM-like
power spectrum, now known empirically to be a good approximation. 

\citet{PFleury14a}
demonstrated with completely analytic arguments the equivalence of the
ZKDR distance and that calculated from a certain class of Swiss-cheese
models at a well controlled level of approximation.  This had been known
for a long time based on comparisons of numerical results, but of course
an analytic proof is very important.  Since the Swiss-cheese models are
exact solutions of the Einstein equations, this means that there can be
no problem using the ZKDR distance, as long as one makes the reasonable
assumption that the mass at the centre of a Swiss-cheese hole is
effectively opaque and reasonable assumptions about the order of
magnitude of the mass and compactness of the clumps.  (Of course, as
discussed in \Sect{mzr}, even if there can be no debate that the ZKDR
distance is appropriate if a \universe{} has the corresponding mass
distribution, it is another question whether our \Universe{} does indeed
have such a mass distribution, even approximately.)  He also stressed
that the Etherington reciprocity relation (\eq{Etherington}) holds for
\emph{any} \spacetime{} in which the number of photons is conserved, a
point which is sometimes misunderstood.  The present work is concerned
with the theory and applications of the ZKDR distance, assuming that it
is correct. 
\citet{PFleury14a}
has written the definitive paper on the justification of the ZKDR
distance; it and references therein should be consulted for those
interested in details. 

\citet{APeelTI14a,APeelTI15a}
examined the effcts of inhomogeneities on distance measures in a
Swiss-cheese model, concentrating on the distance modulus.  Their model
is more general because the holes are non-symmetric structures described
by the 
\citet{PSzekeres75a}
metric (in general inhomogeneous and anisotropic).  This allows an exact
description which includes non-trivial evolution of structure.
Interestingly, the standard deviation for dispersions $\Delta\mu$ was
found to be $0.004\le\sigma_{\Delta\mu}\le0.008$, smaller than the
intrinsic dispersion of magnitudes of \tIas{}. 

\citet{MLavintoSRasanen15a}
examined the CMB as seen through random Swiss cheese.  Usually, `closed'
holes had been examined, \ie{}~an overdense centre surrounded by an
underdensity. 
\citet{MLavintoSRasanen15a}
examind `open' holes as well, \ie{}~an underdense void surrounded by a
thin overdense shell.  This is arguably a better model of our
\Universe{}, though of course still an approximation.  The size of the
holes corresponds to galaxy clusters.  There is no statistically
significant systematic shift in the angular-diameter distance, with a
95\percentadj{} upper limit of $|\Delta \Dasd{}/\bar{\Dasd{}}|<10^{-4}$,
and larger values reported in the literature are shown to be due to
selection effects. 

Observed inhomogeneities in the CMB are caused by a combination of
primordial inhomogeneities and the effects of inhomogeneities on light
propagation.  Since the relevant angular scales are much larger than
those involved in the ZKDR distance, further discussion of CMB
anisotropies is beyond the scope of the present work. 
\citet{MLavintoSRasanen15a},
apart from presenting original results, also gave a good review of this
topic and its connection to the ZKDR distance.

\section{Gravitational lensing: time delays}
\label{time-delays}

The basic observational quantities in a strong (\eg{}~multiple-image)
gravitational lens system\pdash{}angles, flux ratios\pdash{}are
dimensionless, except for the time delays between pairs of images 
\citep{SRefsdal64b}.
This allows one to determine the Hubble constant from a measurement of
the time delay, assuming a mass model for the lens.  However, this is
true only in the low-redshift limit; at higher redshift, the
cosmological model plays a role 
\citep{SRefsdal66a}.
The cosmological parameters \onull{} and \lnull{} are now known very
well from cosmological tests other than gravitational-lensing time
delays 
\citep[\eg{}][]{PLANCKXIV14a,PLANCKXIII2016a,PLANCKVI2019a}; 
one could thus assume them to be exactly known and use observations
related to cosmological distances to determine $\eta$ 
\citep[\eg{}][]{PHelbig15Ra}.%
\footnote{The data from these other tests cannot usefully constrain
\onull{}, \lnull{}, and $\eta$ simultaneously 
\citep{VBustiSL12a,PHelbig15Ra}, 
not even if one restricts the analysis to a flat \universe{}; the same
is true of similar tests involving the \asrr{} 
\citep{RSantosJLima08a}.%
}
Within the uncertainties as they were 35\rdash{}40 years ago, for the
\asd{}, at low redshift the values of \onull{} and \lnull{} are more
important, while $\eta$ becomes more important at high redshift
(\eg{}~\figr{1} in KHS).  Due to the different combination of \asd{}s,
for lensing statistics the effect of $\eta$ tends to cancel 
\citep[\eg{}][]{RQuastPHelbig99Ra}
while in the case of gravitational-lensing time delays the importance of
$\eta$ is enhanced even at lower redshift 
\citep[\eg{}][]{RKayserSRefsdal83a,PHelbig97Pa}.

\citet{RKayserSRefsdal83a}
illustrated this dramatically for several world models with
$\lnull{}=0$, comparing the $\eta=1$ and $\eta=0$ cases.  For the double
quasar 0957+561 
\citep{DWalshCW79a},
the cosmological correction factor (which gives the influence of the
cosmological model compared to the limiting low-redshift case) was
calculated for $\sigma_{0}$ values ranging from 0 to 2 (corresponding to
$0\le\onull{}\le4$) with $\qnull{}$ values of 1.0, 0.5, 0.0, and $-1$
($\lnull{}=\snull{}-\qnull{}$). 
\citet{PHelbig97Pa}
repeated the exercise for arbitrary combinations of \lnull{}, \onull{},
and $\eta$, again showing the importance of $\eta$, which has become
even more important now that the values of \lnull{} and \onull{} are so
well known. 

A somewhat more complicated model (not neglecting shear) was
investigated by 
\citet{CAlcockNAnderson85a},
for $\lnull{}=0$ (not stated but assumed) and \onull{} values of 0 and
1, using two gravitational-lens systems as concrete examples.  They
stressed the fact that ignorance of the mass distribution along the line
of sight makes it difficult to determine the Hubble constant by this
method, but also that, once the Hubble constant is known \via{} other
means, this method could be used to learn something about the mass
distribution.  Similar results were obtained by 
\citet{KWatanabeST92a}.

Usually one thinks of the possibility of determining \hnull{} or, if
\hnull{} is known, other cosmological parameters from a measured time
delay and mass model for the lens. 
\citet{RNarayan91a}
pointed out that the measurement actually gives one the \asd{} between
observer and lens (which, if the redshift of the lens is known, is
easily converted into the Hubble constant).  Of course, this depends on
$\eta$, but since lens redshifts are usually low, the effect of $\eta$
is limited. 

\citet{FGioviLAmendola01a}
examined a more general quintessence model where, in addition to
ordinary matter (`dust') there is a perfect fluid with equation of state
$p = (\frac{m}{3}-1)\rho$ with $0 \le m<3$.  The case $m=0$ corresponds
to the cosmological constant while $m=3$ corresponds to ordinary matter;
$m<2$ implies that the \universe{} is accelerating (as long as the
quintessence term dominates).  However, only $k=0$ models are
considered.  One might think that this is justified since the
\Universe{} does seem to be very close to being flat 
\citep[\eg{}][]{PLANCKXIV14a,PLANCKXIII2016a,PLANCKVI2019a}; 
however, such an interpretation usually assumes that $m=0$.
Nevertheless, all known analytic solutions within this framework are
presented (except one which `is so complicated that it is not worth
reporting').  Other cases are calculated numerically.  Including
quintessence usually reduces the estimated value of \hnull{} compared to
the standard $m=0$ case.  Marginalizing over \onull{} and $m$ for the
time delays considered results in $\hnull{}=71\pm6$ and
$\hnull{}=64\pm4$ km/s/Mpc for the cases $\eta=0$ and $\eta=1$,
respectively.  Considering the facts that there is no evidence at all
for values of $m$ other than 0 (the cosmological constant) and 3 (dust),
apart from radiation with $m=4$ which, however, is important only in the
early \Universe{}, and that $\eta=1$ is obviously not correct (at least
in the strict sense), I find it somewhat disconcerting that there are a
large number of papers investigating the possible effects of
quintessence on the interpretation of cosmological observations compared
to the number which discuss the influence of $\eta$. 

While the idea is simple in principle
\citep{SRefsdal64b},
in practice many details need to be taken into account when determining
\hnull{} from gravitational-lens time delays (especially if the
uncertainties should be small enough to be competitive with other
methods), such as measuring the time delay itself and determining
realistic uncertainties 
\citep[\eg{}][]{ABiggsIBrowne18a}
and constructing a realistic mass model for the lens
\citep[\eg{}][]{KWongetal16a,CRusuetal19a}.  At this level of detail,
characterizing the density along the line of sight by a single parameter
$\eta$, or even $\eta(z)$, is too coarse.  Rather, one attempts to
measure the mass distribution explicitly, by counting galaxies 
\citep[\eg{}][]{CRusuetal17a}
or using weak gravitational lensing
\citep[\eg{}][]{OTihhonovaetal18a}.

\section{Gravitational lensing: amplification}
\label{amplification}

\citet{PSchneider84a} 
showed that a general transparent mass distribution always leads to
amplification of at least one image compared to the case of an $\eta=0$
\universe{} (\ie{}~compared to the case that the lens were absent, not
compared to the case that its mass is smoothly distributed throughout
the \universe{}).  Of course, this is not in contradiction with the
result of 
\citet{SWeinberg76a}
that there is no mean amplification compared to a \emph{homogeneous}
\universe{}, a point also emphasized by 
\citet[][see \Sect{S-c}]{LNottaleFHammer84a}
and
\citet{FHammer85a}.

Of course, all discussion of the ZKDR distance involves (negative)
amplification, and in general all gravitational lensing involves
amplification.  Gravitational lensing has a huge literature which is
beyond the scope of the present work.  Therefore, I discuss here only
those aspects of gravitational lensing which are directly related to the
ZKDR distance, are interesting for other reasons, or in which I was
personally involved.  One example of the last is a study 
\citep{EZackrissonetal03Ra}
which demonstrated that various claims
\citep{MHawkins93a,MHawkins96a,MHawkins97b,MHawkinsNTaylor97a}
that most dark matter must be in compact objects of about a solar
mass\pdash{}because this is assumed to be responsible for most of the
long-term optical variability of QSOs \via{} microlensing\pdash{}cannot
be correct.  In short, while arguments were presented that many of the
observations are not only compatible with microlensing but also have no
other obvious explanation, there are nevertheless other observations
which contradict this hyposthesis, in particular the distribution of
amplifications.

\section{Gravitational lensing: general}
\label{general}

\citet{CAlcockNAnderson86}
qualitatively discussed the optical scalars\pdash{}implying a model more
complicated than the ZKDR distance\pdash{}and the possibility to learn
something about distribution of mass in the \universe{} from the
distance measures derived from gravitational-lens systems.  (Often the
reverse is done: one has some model to calculate the distance as a
function of redshift, and uses this as input for modelling the lens
system.)  Perhaps because in the case of gravitational lensing it is
obvious that there are small-scale inhomogeneities which affect light
rays (\ie{}~the gravitational lenses themselves), the ZKDR distance and
similar topics were discussed earlier and more often than in other
areas, even though their role there could be just as important. 

\citet{MLeeBPaczynski90a}
investigated gravitational lensing by three-dimensional mass
distributions, finding that 16 screens are a sufficiently good
approximation.  Their conclusion that `the distribution of
amplifications of single images is dominated by the convergence due to
matter within the beam' and that `[t]he shear caused by matter outside
the beam has no significant effect'\pdash{}even in the case of strong
lensing\pdash{}increases one's confidence that the zero-shear ZKDR
distance is a realistic approximation (at least in a \universe{} with
the corresponding mass distribution).  Although their goal was not to
test the ZKDR approximation, their work could be seen as an early
comparison of the ZKDR distance with numerical simulations. 
\citet{MJaroszynskiPPG90a}
numerically studied gravitational lensing in the \EdS{} model, also
concluding that shear can be neglected but also that the filled-beam
approximation ($\eta=1$) appears to be justified, at least for strong
lensing by galaxies or clusters of galaxies.  However, `the column
density was averaged over a comoving area of approximately $(1h^{-1}
\mathrm{Mpc})^{2}$', so this could be a self-fulfilling prophecy,
together with the fact that they found no case of strong lensing.
Nevertheless, it does seem to be the fact that `the large-scale
structure of the universe as it is presently known does not produce
multiple images with gravitational lensing on a scale larger than
clusters of galaxies'.  The same conclusion, namely that Weyl focussing
can be neglected compare to Ricci focussing, was also found by 
\citet{THamana99a}
to apply to a \universe{} modelled as randomly distributed isothermal
objects.  It thus appears that the ZKDR distance, which is based on a
very simple model, is also valid in more-realistic models, confirming a
result of 
\citet{TNakamura97a}
based on solving the optical-scalar equation for light passing through
linear inhomogeneities in CDM models. 

\citet{SSeitzSE94a}
and
\citet{SSeitzPSchneider94a}
derived the gravitational-lens equations in an `on average' Friedmann
\universe{}, in particular one with the mass distribution (smooth
component with clumps) used in the derivation of the ZKDR distance. 
This very detailed work is an analytic complement to the numerical
investigations mentioned above regarding the effects of inhomogeneities
on the propagation of light beams; in particular, necessary
approximations are made clear, lending support to the idea that the ZKDR
distance is an acceptable approximation. 

Gravitational-lensing statistics
\citep[\eg{}][]{ETurnerOG84a,MFukugitaFK90a,MFukugitaFKT92a,%
EFalcoKM98a,CKochanek93c,CKochanek96a,CKochanek96b,CKochanekFS95a,%
RQuastPHelbig99Ra,PHelbigetal99Ra,KHChaeetal02a}
is usually not concerned with $\eta$.  Apart from the general neglect of
$\eta$ in observational cosmology, there are probably several reasons
for this.  First, such studies are usually concerned with all-sky
surveys, so one might expect $\eta$ to `average out' to 1 
\citep{SWeinberg76a}.  
Second, in the relevant combination of \asd{}s, the effect of $\eta$
tends to cancel out (in contrast to the situation regarding time
delays).  Third, while selection effects are important in such analyses,
selection effects due to the value of $\eta$ are smaller than others.
Fourth, any effect of $\eta$ would, in practice, be degenerate with
other effects. 
\citet{GCovoneSdR05a}
found that the expected number of gravitationally lensed quasars is a
decreasing function of $\eta$; 
\citet{LCAstanedaDValencia08a}
investigated strong lensing (by galaxy clusters) with $\eta=\eta(z)$ as
a means of taking structure formation into account.%
\footnote{Note that one expects $\eta$ to increase with $z$ for two
reasons when the \asd{} is concerned.  First, structure formation
implies that the \universe{} is more homogeneous at higher redshift.
Second, for a fixed angle at the observer, the physical size of the
object observed increases with redshift (as long as the redshift is
lower than that of the maximum in the \asd{}), so one averages over a
larger volume at higher redshift.  Both effects exist for the luminosity
distance as well.%
} 

\citet{HAsada98a},
by contrast, assumed the validity of the ZKDR distance and used it to
investigate how inhomogeneities affect observations of gravitational
lenses, in particular bending angle, lensing statistics, and time delay.
An interesting analytic result is that all three combinations of
distances%
\footnote{The combinations are \Dds{}/\Ds{}, \Dd{}\Dds{}/\Ds{}, and
\Dd{}\Ds{}/\Dds{}, respectively.  The subscripts refer to the deflector
(lens) and source.  In the case of only one subscript, it is the second,
the first being understood to refer to the observer.  This is probably
the most common notation.  Other schemes explicitly write the first
subscript when it refers to the observer as well, use `l' instead of `d'
to refer to the lens (deflector), use capital letters, or some
combination of these.  The same subscripts are used to refer to the
corresponding redshifts, \eg{}~\zs{}, though sometimes \zd{} is used in
the sense of a variable and \zl{} to refer to the redshift of an
explicit gravitational lens.%
}
involved in these phenomena are monotonic with respect to the clumpiness
for all combinations of \lnull{}, \onull{}, and source and lens
redshifts.  The clumpiness decreases the bending angle and number of
strong-lensing events and increases the time delay.  (Of course, not all
combinations are monotonic in $\eta$, but physically relevant ones are.)
In the first two cases, decreasing $\eta$ has the same effect as
decreasing \lnull{}.  In other words, using a value of $\eta$ which is
too large (such as the common assumption $\eta=1$) would lead one
underestimate the value of \lnull{}.%
\footnote{Note that this is opposite the effect in the \mzr{}.%
}
(In the conclusions, this is confusingly stated as `the use of the DR
distance always leads to the \emph{overestimate} of the cosmological
constant' [emphasis in the original]; of course, it is not an
overestimate but rather the correct estimate if the correct value of
$\eta$ for the ZKDR distance is used.)  More detail was provided by 
\citet{KTomitaAH99a}. 

At almost the same time (publication was one month later) and completely
independently, 
\citet{PHelbig98Ra}
investigated not the common gravitational-lensing topics mentioned
above, but rather the correlation between image separation and source
redshift, in a reply to the work of 
\citet{MParkJGott97a}
who had noted a negative correlation. 
\citet{PHelbig98Ra}
showed that decreasing $\eta$ has the same effect as decreasing
$K:=\lnull{}+\onull{}-1$ (\ie{}~this effect is also monotonic in
$\eta$); also, decreasing $\eta$ reduces the differences between
cosmological models characterized by \lnull{} and \onull{}.  The strong
negative correlation reported by 
\citet{MParkJGott97a},
though, seems to be based on an unclean data sample and also is not
statistically significant.

It had been known for some time
\citep[\eg{}][]{PSchneiderEF92a,JEhlersPSchneider86a}
that gravitational-lensing magnification as calculated using the
standard distance is smaller than that using the ZKDR distance by a
factor of the square of the ratio of the corresponding distances, a
result derived by averaging magnifications over a number of sources and
making use of flux conservation. 
\citet{THamana98a}
showed that it is actually true not just on average but for each
individual ray bundle as well.

\section{Monte-Carlo simulations}

\citet{SRefsdal70a}
had studied numerically the propagation of light in an inhomogeneous
\universe{} (see \sect{other1}).  This technique was expanded by 
\citet{PSchneiderAWeiss88a,PSchneiderAWeiss88b}.
\citet{YPei93a,YPei93b}
showed that, to a reasonable approximation, the effect of multiple
lenses can be calculated by multiplying the individual amplifications. 
\citet{KTomita98a}
used N-body simulations with the CDM power spectrum in four cosmological
models to investigate the behaviour of angular-diameter distances in
inhomogeneous cosmological models, determining $\eta$ for each pair of
rays and investigating the mean and dispersion of $\eta$.  Further
studies along these lines 
\citep[\eg{}][]{PPremadiMM98a,PPremadietal01a,HMartelPM02a,PPremadiMM04a,%
PPremadiMM08a}
involving ray shooting through N-body simulations with the explicit
calculation of the paths of (bundles of) light rays, while interesting,
are too far removed from the main topic of the present article for
further discussion.

\section{Classical cosmology: redshift\jdash{}volume relation}
\label{volume}

\citet{MOmoteHYoshida90a}
examined the effect of statistical gravitational amplification on the
cosmological redshift\jdash{}volume test, in particular its influence on
the derived value of \onull{}, using the extreme $\eta=0$ model to
examine the data of 
\citet{ELohESpillar86a},
concluding that their derived value of \onull{} is smaller, \ie{}~$\eta$
and \onull{} are positively correlated.%
\footnote{Note that in the simpler case of the \mzr{}, $\eta$ and
\onull{} are negatively correlated.  This is easy to understand, since
both a higher value of $\eta$ and a higher value of \onull{} mean that
more matter is in the beam.  In the redshift\jdash{}volume test, both
the apparent magnitude and the volume (which is independent of $\eta$)
are involved, the luminosity function plays a role, \etc{}, making the
test much more complicated;
\citep[\eg{}][]{ASandage95a}.
Also, all mass was assumed to be in point masses with regard to the
gravitational-lens effect. 
\citet{HYoshidaMOmote1992a}
performed a similar study using the model of a spherical opaque lens,
arriving at similar conclusions.%
}
Of course, there are much better data today, 
\citet{ELohESpillar86a}
neglected galaxy evolution, and so on; nevertheless, this work
demonstrates the effect of $\eta$ on the redshift\jdash{}volume test.

\section{Classical cosmology: magnitudes}

\citet{XWU90a}
suggested that interest in the ZKDR distance had subsided after 
\citet{SWeinberg76a}
had shown that flux conservation implies that, on average, there is no
amplification.%
\footnote{This is not my impression.  There was a slow trickle of papers
up until about 1982, after which the number per year increased each
year.  This appears to be mainly data-driven, with a large increase
after the measurement of the \mzr{} for \tIas{}.%
}
He then points out that the fact that the luminosity distances in the
homogeneous and inhomogeneous cases are the same on average does not
mean that apparent magnitudes are the same in both cases.  This is
illustrated with a simple model.  More important than the model are the
conclusions: because most lines of sight are underdense, compensated by
the occasional large amplification, the apparent magnitude is
essentially a random variable; also, the value of \qnull{} obtained
depends on the value of $\eta$ assumed, or, \viceversa{}, one could use
the \mzr{} to determine $\eta$ if the cosmological parameters are known
with some degree of certainty. 

\citet{HRose01a}
pointed out that the argument of 
\citet{SWeinberg76a}
does not hold if the sphere centred on the observer is affected by the
mass distribution, concluding that, in a perturbed \FRW{} \universe{},
`more photons from a source at a given redshift' will be received than
in \aFRW{} \universe{}, \ie{}~the sources are brighter.  Somewhat
confusingly, it is claimed that they `therefore have a higher apparent
magnitude', which is correct if `higher' means `brighter', but of course
larger magnitudes correspond to fainter objects.  However, this is a
second-order effect; to first order, small deviations from homogeneity
do not change the average magnification 
\citep{CClaudel00a}.

\section{Classical cosmology: magnitude\jdash{}number relation}

Although going somewhat beyond the simple approximation of the ZKDR
distance, 
\citet{KWatanabe92a,KWatanabe93a}
investigated the effects of an inhomogeneous \universe{} on another
classic cosmological test, namely the magnitude\jdash{}number relation 
\citep[see \eg{}~][for details]{ASandage95a}, 
also checking the validity of the assumptions used by 
\citet{MOmoteHYoshida90a}
(see \sect{volume}).  These sorts of cosmological tests have gone out of
fashion, primarily because the uncertainty in the evolution of the
sources is too large, leaving the \mzr{} for \tIas{}, baryon acoustic
oscillations (BAO), and the CMB as the most useful cosmological tests.
It is not yet possible to calculate galaxy evolution from first
principles, and observations of it have to be interpreted within the
context of an assumed cosmological model, so now such classic tests are
useful mainly as consistency checks.

\section{Classical cosmology: magnitude\jdash{}redshift relation}
\label{mzr}

One of the most important advances in observational cosmology has been
the application of the \mzr{} to \tIas{}.%
\footnote{The \mzr{} for \tIas{} has spawned an extensive literature; in
this review, I mention only those aspects of it directly concerned with
the ZKDR distance.  Many good reviews are available 
\citep{ARiess2000a,BLeibundgut01a,BSchmidt02a,SPerlmutterBSchmidt03a,%
AFilippenko05a,BLeibundgut2008a}.%
} 
In an influential paper,
\citet{SColgate79a}
had suggested using the Hubble Space Telescope for that purpose. 
\citet{AGoobarSPerlmutter95a}
discussed the feasability of such a programme, and were later involved
in the Supernova Cosmology Project, which reported measurements of
\lnull{} and \onull{} based on 42 supernovae 
\citep{SPerlmutteretal99a,RKnopetal03a}, 
a result confirmed and published slightly earlier by the High-$z$
Supernova Search team 
\citep{ARiessetal98a,BSchmidtetal98a}.
While there had been hints, based on joint constraints from several
cosmological tests, not only that the cosmological constant is positive
but also that it has such a value that the \Universe{} is currently
accelerating 
\citep{JOstrikerPSteinhardt95a,KKraussMTurner95a},
the \mzr{} for \tIas{} was the first cosmological test which, by itself,
confirmed such a value for \lnull{}.  (Contrary to some claims, this
test does not `directly' measure acceleration in any meaningful sense,
even if one does not adopt the extreme view that all that is ever
`really' measured in observational astronomy, whether in imaging or in
spectroscopy, are photon counts as a function of position on a
detector.) 
\citet{SPerlmutteretal99a}
also checked for the influence of $\eta$, using the \code{Fortran} code
of KHS to compare the standard distance to that of two other models, one
with $\eta=0$ and the other with $\eta=\eta(\onull{})$, the latter based
on the idea that all matter is in clumps for $\onull{}\leq0.25$ and for
$\onull{}\ge0.25$ the fraction 0.25/\onull{} is in clumps, thus $\eta=0$
for $\onull{}\leq0.25$, otherwise $\eta=1-0.25/\onull{}$.  Their
conclusion, based of course on their data at the time, is that
significant differences occur only for models ruled out by other
arguments, \ie{}~$\onull{}>1$. 

\citet{RKantowskiVB95a},
still using the soon-to-be-obsolete $\qnull{}$-notation, had pointed out
that $\eta$ should be taken into account when discussing the \mzr{} for
\tIas{}.  They also presented an analytic solution for $\lnull{}=0$ but
arbitrary \onull{} and \qnull{}, and introduced the parameter $\nu$: 
\begin{equation}
\label{nu}
\eta = 1 - \frac{\nu(\nu+1)}{6} \mathspace{} ,
\end{equation}
due to the fact that there are analytic solutions for certain integer
values of $\nu$. 
\citet{JFRieman97a}
disputed the importance of the effect, arguing that the Swiss-cheese
model is not a valid model for the distribution of mass in the
\Universe{}, and that the uncertainty due to $\eta$ would be smaller;
\citet{RKantowskiVB95a} disagree.
\citet{JFRieman97a}
emphasized the dispersion in the apparent magnitude of supernovae caused
by a given mass distribution, rather than considering a range of $\eta$.
A similar approach, with the aim of determining the density of compact
objects, the properties of galaxy haloes, or estimating the uncertainty
in the measurement of \lnull{} and \onull{}, was taken up by many
authors 
\citep[\eg{}][]{DHolz98a,USeljakDHolz99a,RMetcalfJSilk99a,%
PValageas00a,EMoertsellGB01a,EMintyHH02a,RAmanullaMG2003a,%
TPayneMBirkinshaw04a,RMetcalfJSilk07a,SDodelsonAVallinotto06a,%
HMartelPPremadi08a,CYooetal08a,JJoenssonetal10a,%
IBen-DayanRTakahashi16a,MZumalacarreguiUSeljak18a}.
Rather than calculating the dispersion, one could also attempt to
measure it indirectly due to the fact that the same matter fluctuations
would cause weak lensing.  However, the shear maps smoothed on
\arcminutea{} scales are not of much use since an appreciable fraction
of the lensing dispersion derives from \subarcminutea{} scales 
\citep{NDalaletal03a}.
Another approach is to estimate the amplification from the matter
visible along the line of sight; 
\citet{JJoenssonetal06a,JJoenssonetal07a,JJoenssonetal08a}
and
\citet{MSmithetal14a},
building on ideas by
\citet{CGunnarssonetal06a},
found a tentative detection, \ie{}~a correlation between the computed
and observed amplification (difference between the observed flux and
that expected from the redshift in the concordance model).  One can also
turn this around, and use the observed matter distribution to estimate
the amplification due to lensing and thus correct the observed flux
\citep{JJoenssonMS09a}. 

\citet{KIwataCYoo15a}
took a somewhat different approach, assuming a flat \universe{} and
taking \onull{} from CMB measurements, then calculating $\eta(z)$ such
that the cosmological parameters from the \mzr{} for \tIas{} agree; this
was done for four different scenarios.  This is complementary to the
work of 
\citet{PHelbig15Ra}
(next paragraph) who, at almost exactly the same time, considered only
constant $\eta$ but for arbitrary \FRW{} models, determining the value
of $\eta$ such that the \mzr{} for \tIas{} results in the same values
for \lnull{} and \onull{} as those derived from the CMB. 

\citet{PHelbig15Ra}
investigated the influence of $\eta$, noting that more and
higher-redshift data had become available.  While the data were not good
enough to determine \lnull{}, \onull{}, and $\eta$ simultaneously%
\footnote{This would imply the somewhat dubious assumption that $\eta$
is independent of both redshift and the line of sight.  Of course,
more-realistic models could take such effects into account, but
obviously the data would not be able to constrain them since even the
simpler model with a constant $\eta$ could not be constrained.%
}%
, 
the constraints in the \lnull{}\jdash{}\onull{} plane depend strongly on
$\eta$.  Only by assuming $\eta\approx1$ does one recover the
concordance-cosmology values of $\lnull{}\approx0.7$ and
$\onull{}\approx0.3$.  Since these values are now known to high
precision independently of the \mzr{} for \tIas{} 
\citep[\eg{}][]{PLANCKXIV14a,PLANCKXIII2016a,PLANCKVI2019a},
one can use the \mzr{} for \tIas{} to measure $\eta$.  The result
$\eta\approx1$ agrees well with other tests to determine $\eta$ from
observations.  (While no useful constraints are possible, the global
maximum likelihood in the \lnull{}\jdash{}\onull{}\jdash{}$\eta$ cube
also indicates a high value of $\eta$.)  Unknown to me at the time, very
similar results, based on the same data, were obtained by 
\citet{XYangYZ13a},
\citet{NBretonAMontiel13a},
and, somewhat later,
\citet{ZLiFZ15a}
(the latter two restricted to a flat \universe{}).  While perhaps not
surprising, it is of course important in science for results to be
confirmed by others working independently.  Although they investigated a
wider range of models, when restricted to standard \FRW{} models, the
results of 
\citet{SDhawanGM18a}
are also consistent.

Since the observations indicate that $\eta\approx1$, one can ask whether
this is true `on average' as discussed by 
\citet{SWeinberg76a}, 
or whether each line of sight indicates $\eta\approx1$.  In the former
case, one would expect a dispersion in the distance at high redshift.
Indeed, the scatter does increase with redshift, but so do the
observational uncertainties.  Since their quotient is independent of
redshift, this indicates that each line of sight indicates
$\eta\approx1$, in other words that all lines of sight fairly sample the
mass distribution of the \Universe{}%
\footnote{As discussed in \sects{flux1}{\rdash{}}{S-c}, this does not
imply that the \Universe{} is effectively homogeneous, but rather that
the distance calculated from redshift is approximately the same as that
which would be calculated in an effectively homogeneous \universe{}.%
}
\citep{PHelbig15Rb}.
Note that 
\citet{DHolzELinder05a}
find a scatter (calculated theoretically) approximated by a Gaussian
with standard deviation $\sigma_{\mathrm{eff}}=0.088z$ (in flux) or
$\sigma_{\mathrm{eff,m}}=0.093z$ (in magnitudes).  However, as
discussed by 
\citet{PHelbig15Rb},
the observed increase in scatter with redshift seems primarily due to
observational uncertainties in addition to the theoretically calculated
scatter sometimes incorporated into those uncertainties.

\section{More-detailed models}
\label{more-detailed}

\citet{DHolzRWald98a}
developed a generalization of the Swiss-cheese approximation by
including all mass explicitly (thus there is no smoothed-out `cheese'
component), requiring the mass within a given spherical region
(corresponding to a hole in the Swiss-cheese approach) to be equal to
that of the background \FRW{} model only on average, and dropping the
requirement of spherical symmetry.  In addition, rather than having a
fixed mass distribution and calculating the trajectories of photons
within it, the mass distribution along a given trajectory is calculated
on the fly.  Also, no opaque-radius cutoff is imposed.  Such a model is
clearly more realistic than that of Zel'dovich or a Swiss-cheese model,
and leads to a distribution of apparent luminosities at a given
redshift.  In principle, the shape of such a distribution can be used to
determine both the background \FRW{} model and the fraction of matter in
compact objects.  While there are a few highly amplified sources (which,
due to flux conservation, there must be, in order to compensate for the
fact that most sources are de-amplified), most of the distribution can
be thought of as $\eta$ varying with position on the sky.  As expected,
if thought of in terms of $\eta$, $\eta$ increases with redshift, as the
higher the redshift, the more likely it is that a typical trajectory
crosses a fair sample of the \universe{}. 

\citet{LBergstroemGGM2000a}
generalized the method of
\citet{DHolzRWald98a}
by allowing for different types of fluids, possibly with non-vanishing
pressure, instead of just dust, and by considering the NFW profile 
\citep{JNavarroFW97a}
in addition to point masses and singular isothermal spheres as lenses 
\citep[see also][]{MGoliathEMoertselll00a}.  
Also, multiple imaging is taken into account.  This is thus an even more
complicated and thus more realistic model of the \universe{}.  As a
consistency check, their results for empty cells and cells with a
homogeneous dust component were compared with results obtained from the
code of KHS for $\eta=0$ and $\eta=1$, respectively.  For a variety of
cosmological models, the discrepancy was less than 1\percent{} up to
$z=10$.  This is a further justification that the ZKDR distance is an
excellent approximation provided that the mass in the \universe{} is
distributed according to the assumptions underlying the ZKDR distance.
They also found analytic approximations which are very good
representations of various observable quantities, such as magnification
distributions. 

\citet{EMoertsell2002a}
used essentially the same scheme to investigate the relation between
$\eta$ and the fraction of compact objects.  By definition, $1-\eta$ is
the fraction of compact objects $f_{\mathrm{c}}$ in the pure ZKDR case,
\ie{}~only de-amplification due to underdensity and no amplification due
to gravitational lensing.  As expected, taking lensing into account
results in $1-\eta < f_{\mathrm{c}}$.  Interestingly, for a variety of
cosmological models ($(\onull{},\lnull{})=(0.3,0.6), (0.2,0.0),
(1.0,0.0))$, for redshifts between 0 and 3, and for various models of
the mass distribution (homogeneous and point masses, NFW profiles and
point masses), the relation is approximated very well by
$1-\eta\approx{}0.6f_{\mathrm{c}}$. 

Some authors have claimed that that a \universe{} with
\emph{large-scale} inhomogeneities could appear as if it has a positive
cosmological constant when in fact it doesn't, either because the \mzr{}
mimics that of an accelerating model 
\citep[\eg{}][]{HAlnesAG06a,DGarfinkle06a}
and/or because the inhomogeneities produce accelerations without a
cosmological constant 
\citep[\eg{}][]{TKaietal07a}.
However,
\citet{RVanderveldFN06a}
present evidence against these claims.  Also, while in principle one can
reproduce an arbitrary \mzr{} with an \adhoc{} mass distribution, there
are two arguments against this, other than the fact that it is
\adhoc{}\pdash{}or, equivalently, of all possible \mzr{} which could be
produced, it just so happens that one is produced which is not only
explicable with 1920s cosmology, but also where the derived parameters
agree with those determined by other means\pdash{}: there is no
believable route to explaining the CMB observations, and we are required
to be at or near the centre of a large and approximately spherical
region.  Those topics go beyond the scope of this article, so I don't
discuss them further here.  However, it has even been claimed that this
is possible in a Swiss-cheese \universe{} 
\citep[\eg{}][]{VMarraetal07a,VMarraKM8a}.
\citet{RVanderveldFW08a}
showed, however, that this is not the case if the voids have a random
distribution. 

\citet{EFlanaganetal12a}
used a variant of the method of
\citet{DHolzRWald98a}
to calculate the distribution of magnitude shifts, but using a
simplified Swiss-cheese model for the mass distribution. 
\citet{EFlanaganetal13a}
extended this with a more refined Swiss-cheese model: the mass removed to
make the voids is distributed on shells surrounding the holes in the
form of randomly located NFW \haloes{} and in the interior of the holes
(either smoothly distributed or as randomly located \haloes{}). 

\citet{RHadaTFutamase14a}
carried out a similar exercise, concentrating on the difference between
the magnitude\jdash{}redshift relation in a homogeneous \universe{} and
that in an inhomogeneous \universe{} (with a mass distribution given by
the non-linear matter power spectrum), as well as its dispersion, taking
into account the blocking effect by collapsed objects and examining the
resulting uncertainty in \onull{} ($\approx 0.4$) and the equation of
state $w$ ($\approx 0.04$), all in a flat \universe{}. 

The work by 
\citet{JGiblinMS16a,JGiblinMS16b}
and 
\citet{JMertensGS16a}
has been mentioned above in \Sect{testing}; a similar approach was
adopted by 
\citet{EBentivegnaMBruni16a}.  
Detailed discussion of such work is of course beyond the scope of this
review, which concentrates on the use of the ZKDR distance as opposed to
the standard distance when calculating distance from redshift for a
given cosmological model.  Nevertheless, for present purposes such works
are interesting because they allow for comparison between the ZKDR
distance and much more realistic simulated matter distributions, making
it possible to see how well the ZKDR \ansatz{} approximates reality.
However, such simulations are still not entirely free of approximations:
those above are fully relativistic but use the fluid approximation,
while a different approach was adopted by 
\citet{AAdameketal16a},
which does not rely on the fluid approximation, but on the other hand is
based on a weak-field expansion of GR.  Which approach is better of
course depends on what one wants to study.  It is perhaps surprising
that a simple equation such as \Eq{dgl} agrees so well with results from
numerical ray tracing through \lcdm{} simulations, at least if one
allows the additional freedom of $\eta(z)$ and a certain stochastic
element depending on the individual line of sight
($\eta(\alpha,\delta$)).  Somewhat similarly, the \FRW{} metric was
originally a simplifying assumption, made in order that at least some
results could be obtained with the limited methods of calculation
available at the time.  Now, however, it is an observational fact, as
demonstrated by observations of the CMB and the large-scale structure of
the \Universe{}, that our \Universe{} is in fact very close to \aFRW{}
model 
\citep{SGreenRWald14a}.

\section{Weak gravitational lensing}

Weak gravitational lensing is normally defined as gravitational lensing
without multiple images.  If the source can be resolved, then
information can be gleaned from the distortion of the image.  In such a
case, however, if the source is at a cosmological distance, $\eta\approx
1$ (because the distance implies a large physical extent near the
source, averaging over the matter distribution, and because it appears
that, at large redshift, distances behave as if $\eta\approx1$, as noted
in \Sect{mzr}).  Relevant for the ZKDR distance with respect to weak
lensing is thus weak lensing of point sources.%
\footnote{An example of strong lensing of resolved sources are multiple
images of background galaxies lensed by clusters of galaxies.  An
example of strong lensing of point sources are multiple images of QSOs;
here, $\eta$ can play a role since it influences the distance calculated
from redshift, which in addition to the lens model is important for the
time delay (see \sect{time-delays}).  Microlensing can be thought of as
a combination of weak and strong lensing, depending on the impact
parameter, though since the source is not resolved, one observes only a
change in apparent magnitude due to amplification.%
} 
Some aspects of this are discussed above in \sect{amplification}.  This
section is concerned particularly with weak lensing of standard candles.

\citet{YWang99a}
pointed out that weak lensing leads to a non-Gaussian magnification
distribution of standard candles at a given redshift, due to the fact
that $\eta$ can vary with direction.  One can thus think of our
\Universe{} as a mosaic of cones centred on the observer, each with a
different value of $\eta$, where there is a unique mapping between
$\eta$ and the magnification of a source.  Of course, since the ZKDR
distance depends on \onull{} and \lnull{} as well as $\eta$, different
cosmological models can lead to very different magnification
distributions for the same matter distribution.%
\footnote{Note that her claim that 
\citet{SPerlmutteretal99a}
`assumed a smooth universe' is somewhat misleading.  While they did not
consider a direction-dependent $\eta$, they did compare the extreme
cases of $\eta=1$ and $\eta=0$ as well as the case of an
\onull{}-dependet $\eta$ (\ie{}~galaxies assigned to clumps and the rest
of the matter distributed smoothly, which implies an increase in $\eta$
with increasing \onull{}), in all cases using the code of KHS.%
}
\citet{YWang99a}
derived an approximation for the ZKDR distance (see
\sect{approximations}), and also treated $\eta$ as a function of
position on the sky, \ie{}~different lines of sight can have different
values of $\eta$.  This effective value of $\eta$ depends not only on
the amount of matter in the beam, but also on how it is distributed
(though only the total amount in the beam is considered\pdash{}the
possibility that a significant fraction could be in point masses is not
taken into account).  An approximation to matter distribution at a given
redshift is found \via{} comparison with the results of 
\citet{JWambsganssetal1997a},
who used $\onull{}=0.4$ and $\lnull{}=0.6$.  She then calculated the
distribution of $\eta$ as well as the magnification distribution for
standard candles, both for the same three different redshifts 0.5, 2,
and 5.  Also, for the same matter distribution, the probability of
magnification was calculated for the same three redshifts and three
different cosmological models: (\onull{},\lnull{}) = (1,0), (0.2,0), and
(0.2,0.8). 

\citet{YWangHM02a}
extended this idea to a universal probability-distribution function for
the reduced convergence which can be directly computed from \onull{} and
\lnull{}, well approximated by a three-parameter stretched Gaussian
distribution, where the three parameters depend only on the variance of
the reduced convergence; in other words, all possible weak-lensing
probability distributions can be well approximated by a one-parameter
family, which was normalized \via{} the simulations of 
\citet{JWambsganssetal1997a}.  
The reduced convergence is the same as the direction-dependent $\eta$
used by 
\citet{YWang99a}.
Fitting formulae were presented for thre fiducial cosmological models:
$(\onull{},\lnull{},h,\sigma_{8}) = (1.0,0.0,0.5,0.6),
(0.3,0.7,0.7,0.9), (0.3,0.0,0.7,0.85)$. 

\citet{LWilliamsJSong04a}
took the opposite approach: assuming that the standard distance
($\eta=1$) is correct, they found that bright SNe are preferentially
found behind regions (5\rdash{}15 \arcminw{} in radius) that are
overdense in the foreground due to $z\approx0.1$ galaxies, the
difference between brightest and faintest being about 0.3\rdash{}0.4
\magnitude{}. (In other words, the fact that bright supernovae are
preferentially found behind overdense regions indicates that the
standard distance is incorrect.)  The effect, significant at
$>99$\percent{}, depends on the amount and distribution of matter along
the line of sight to the sources but not on the details of the
galaxy-biasing scheme. 

In a very detailed work, 
\citet{KKainulainenVMarra09a}
studied the effects of weak gravitational lensing caused by a stochastic
distribution of dark-matter \haloes{}, restricted to flat \FRW{} models
and examining those with $\onull{}=0.28$ (close to the current
concordance model) and $\onull{}=1$ (the \EdS{} model) as representative
examples. In particular, they calculated the difference between the
distance in their model and the ZKDR distance for $\eta=0.5$ and
$\eta=0$ for these two models, finding a maximum relative error of only
0.06 for the extreme case of the empty-beam \EdS{} model at $z=1.6$ (the
upper limit of their redshift range).  This is yet another example of
the proof of the validity of the assumptions underlying the ZKDR
distance.  Their main goal was to compute the probability-distribution
function and the most likely value of the lens convergence along
arbitrary photon geodesics as a function of their model parameters.

\section{Classical cosmology: general}

In an interesting but somewhat confusingly written paper,
\citet{Yuetal11a}
use the \mzr{} and the angular-size\jdash{}redshift relation (based on
data from the literature) to determine \onull{} and $\eta$ in flat
cosmological models (and the equation-of-state parameter 
$w$\pdash{}confusingly referred to as $\omega$\pdash{}and $\eta$ for
flat models with $\onull{}=0.28$).  Of course, $H$ is in general a
function of $z$, but this is not something which is measured directly.%
\footnote{There is a range of directness in measurement.  At one level,
all that is ever measured in astronomy is number of photons as a
function of position on a detector, which can be related to apparent
magnitude for a conventional exposure or as the intensity of a spectrum
in the case of spectroscopy.  Everything else is interpretation.
Nevertheless, it makes sense to say that one can directly measure
redshift, magnitude, and angular size, and, one step less concrete, that
one can measure \lnull{} and \onull{} \via{} the derived parameters
(assuming some framework, such as \FRW{}).  Despite some claims to the
contrary, no cosmological test directly measures acceleration; this is
calculated from the cosmological parameters obtained.  Similarly, $H(z)$
is a calculated quantity.%
} 
Rather, $H(z)$ is calculated from the magnitude or angular size.
Although not stated, presumably the reason is to be able to fit to both
data sets simultaneously.  Their results (1-$\sigma$ uncertainties)
$\eta=0.80\plusminus{0.19}{0.2}$ (with no prior on \onull{}) and
$\eta=0.93\plusminus{0.07}{0.19}$ ($\onull{}=0.26\pm0.1$) can be
compared to $\eta=0.75\plusminus{0.15}{0.15}$ ($\lnull{}=0.72$ and
$\onull{}=0.28$, \ie{}~the confordance-model values) obtained by 
\citet{PHelbig15Ra}
using only the \mzr{} for \tIas{} (see \sect{mzr}).  Although not
directly comparable, and keeping in mind that one would expect the
supernova data to indicate a lower value of $\eta$ due to the smaller
beam size, the general trend is clear: observational data indicate a
relatively high value of $\eta$.  There are two possible explanations.
First, this could be the averaging mentioned by 
\citet{SWeinberg76a},
skewed to slightly lower values because of selection effects.  Second,
the physical model on which the ZKDR distance is based is wrong, but in
our \Universe{} the \mzr{} is similar to that in a high-$\eta$ ZKDR
\universe{} 
\citep{APeelTI14a,PHelbig15Rb} 
(see \sect{mzr}).

\citet{VBustiRSantos11a}
pointed out that the procedure used by 
\citet{Yuetal11a}
to calculate $H(z)$ is not consistent, because the equation relating
$H(z)$ and the \asd{} is valid only for $\eta=1$. 
\citet{RSantosCL08a} 
had done a similar analysis to that of
\citet{Yuetal11a}
using supernova data, concluding that $\eta>0.42$ ($2\sigma$).  Adding
the $H(z)$ data used by 
\citet{Yuetal11a}
of course improves the constraints, resulting in $0.66\le\eta\le1.0$
($2\sigma$) with the best fit at $\eta=1$, a broadly similar result.
Note that 
\citet{PHelbig15Ra}
also finds the \bestfit{} value $\eta=1$ if \lnull{} and/or \onull{} are
constrained.  Thus, while 
\citet{Yuetal11a}
did indeed make a mistake, the fact that $\eta\approx1$ means that it
didn't appreciably affect their main result. 

While there is no evidence that our \Universe{} is not well described by
\aFRW{} model, it is important to test for deviations from this
assumption.  One possibility is to test the Copernican Principle by
looking for a redshift dependence of the curvature parameter 
\citep{CClarksonBL08a};
another is to express \onull{} in terms of observable quantities,
resulting in an expression which must hold at all redshifts 
\citep{VSahniSS08a,CZunkelCClarkson08a}.
\citet{VBustiJLima12a}
pointed out that these tests implicitly assume that the \universe{} is
assumed to be homogeneous and isotropic on all scales (in other words,
the `RW' is assumed; the idea is to test the `F' part of \FRW{}), and
showed that using the ZKDR distance leads to false positives for these
tests (\ie{}~the Copernican Principle appears to be violated when in
fact it is not). 
\citet{VBustiJLima12a}
also rewrite the ZKDR equation so that $\eta$ is given as a function of
observable quantities, allowing one to reconstruct $\eta(z)$ from
observations for a general \lcdm{} model.  Such an $\eta(z)$ can also
mimic the behaviour of model with $\eta=1$ but with $w\ne-1$, \ie{}~some
form of dark energy other than a cosmological constant. 

Inhomogeneous cosmological models definitely affect light propagation.
Whether they affect the expansion rate of the \universe{} is still
debated. 
\citet{SGreenRWald14a}
claimed that there is no evidence that \aFRW{} model is not a good
description of the \Universe{} on essentially all scales (except perhaps
the extremely small scales encountered in, for example, the \mzr{} for
\tIas{}).

\section{Classical cosmology: Angular diameters}

One of the basic cosmological tests is the `standard rod' test,
\ie{}~the comparison of the angular size as a function of redshift of an
object of given size to the theoretical expection derived from the
angular-size\jdash{}redshift relation, which in turn depends on the the
cosmological parameters.  (By the same token, the calculation of the
physical size from the observed angular size depends on the cosmological
model, and on $\eta$.)  Although a classic test, no useful constraints
have been derived from it\pdash{}except in the cases of the CMB and BAO,
though here the corresponding physical lengths are so large that the
ZKDR distance plays no role 
\citep[\eg{}][]{ALewisAChallinor06a}%
\pdash{}primarily because of the difficulty in finding a standard rod.
Nevertheless, some progress can be made.  For example, 
\citet{JAlcanizLS04a}, 
assuming a Gaussian prior $\onull{}=0.35\pm0.07$ in a flat \universe{},
found the best fit at $\onull{}=0.35$ and $\eta=0.8$ (consistent with
the results mention in \sect{mzr}). 

\citet{MAraujoWStoeger09a}
point out the interesting, long-known, but generally unappreciated fact
that, for a flat \universe{}, the redshift at which the maximum of the
\asd{} occurs is a direct measure of \lnull{}, independently of
\hnull{}.  For a non-flat \universe{}, knowledge of the redshift of the
maximum and \hnull{} allows one to determine both \onull{} and \lnull{}.
Note, however, that this depends on the assumption that $\eta=1$. 

Also,
\citet{YChenBRatra12a}
examined constraints from the angular sizes of galaxy clusters, both for
general \FRW{} models and for two classes of flat models with different
types of dark energy.  Their conclusion is still valid today: such
constraints are approximately as restrictive as those based on
gammay-ray\jdash{}burst apparent-luminosity data,
strong\jdash{}gravitational-lensing measurements, or the age of the
\Universe{}, but less so than those from BAO or the $m$\jdash{}$z$
relation for \tIas{} (or the CMB).  Nevertheless, as an independent
constraint, the fact that they are compatible with other data
strengthens our confidence in the concordance model.

\section{Analytic approximations}
\label{approximations}

In general, analytic solutions of the ZKDR distance are very
complicated.  Moreover, there are analytic solutions only for special
values of \lnull{}, \onull{}, or $\eta$. 

\citet{YWang99a} 
presented an approximation for the ZKDR distance as a polynomial in
$\eta$ with coefficients which depend on redshift and the cosmological
parameters, the latter via the fact that the coefficients depend on the
distance calculated for given values of \onull{} and \lnull{} for $\eta$
values of 0, 0.5, 1, and 1.5.  Note that $\eta=1.5$ is in conflict with
the assumptions under which the ZKDR distance is derived; nevertheless,
this can be valid from a heuristic point of view 
\citep[\eg{}][]{JLimaBS14a}.  
Of course, $\eta>1$ everywhere is impossible, but could be valid if
$\eta$ depends on the line of sight.  In that case, however, one should
think of it as an average along the line of sight, \ie{}~a particular
line of sight might, by chance, have an above-average amount of matter
along it.  If this were constant, it would imply an extremely long
structure aligned with the line of sight, which would not be compatible
with an approximate \FRW{} model.

\citet{MDemianskietal03a}%
\footnote{See also
\citet{MDemianskietal00a}
which is the precursor, but longer and substantially different in
places.%
}
found `an approximate analytic solution\dots which is simple enough and
sufficiently accurate to be useful in practical applications'.  It is
not clear how useful this is, though.  It was apparently discovered more
or less by accident and has no theoretical basis.  As such, it is not
clear \apriori{} in which cases it is a good approximation, so one needs
to test it against an (at least numerically) exact solution, in which
case one might just as well use the better solution.%
\footnote{Lewis Carroll, in one of his less famous books, describes a
map with a scale of 1:1, but it was easier to just use the real Earth
than the map \citep{LCarroll93a}.%
}

Also, the numerical implementation of KHS is, in most cases, only a
factor of 3 or so slower than the elliptic-integral solution (of course,
one can compare only in those cases where such solutions exist; the
numerical implementation knows no special cases\pdash{}and is valid for
all values of the input parameters, using the same algorithm for
all\pdash{}and the speed depends only weakly on the input parameters),
so there doesn't seem to be a real need for approximate solutions; even
if such an approximation is faster than the elliptic-integral solution
(and valid for all input parameters), the elliptic-integral solution is
`almost analytic' and reasonably fast, so a factor of 3 for a general
and accurate numerical implementation is not a big disadvantage in
practice.  Though restricted to $k=0$ and $\eta=1$, similar remarks
apply to the work of \citet{UPen99a}.

\section{Summary}

The basis of observational cosmology is calculating the dependence of
some observational quantity\pdash{}usually related to some
distance\pdash{}on redshift for a variety of cosmological models, then
determining the corresponding cosmological parameters \via{} finding the
model which gives the best fit to the data.  Small-scale inhomogeneities
can affect the relation between redshift and distance, thus it at least
needs to be investigated whether results depend on the amount of
inhomogeneity. 
\citet{YZeldovich64at}
introduced a simple model for such small-scale inhomogeneities and an
analytic solution (for the \EdS{} model) for the extreme case, namely
that light propagates through completely empty space, all of the matter
being located in clumps outside the beam.  Subsequent work generalized
that model to other cosmological models and/or intermediate degrees of
inhomogeneity (later known as the ZKDR distance, after the initials of
the most influential pioneers), investigated a similar approach
involving so-called Swiss-cheese models (not necessarily more realistic,
but exact solutions of the Einstein equations) which were later shown to
correspond to the 
\citet{YZeldovich64at}
model in a well defined way, investigated assumptions in the models and
their effects (\eg{}~whether the clumps are transparent, if averaging
whether the average is taken over the celestial sphere or over all lines
of sight, \etc{}), compared the results of the models with exact
solutions or numerical simulations, and developed approximations to
various distance formulae.  Approximations are no longer needed, now
that computing power has increased and an efficient numerical
implementation is available for the general case 
\citep{RKayserHS97Ra}. 

Most of the theory was complete by the middle of the 1970s.  The
discovery of the first gravitational-lens system in 1979 revived
interest in this topic: since gravitational lenses obviously require an
inhomogeneous \universe{}, in such cases the assusmption of a completely
homogeneous \universe{} with regard to light propagation becomes more
obvious.  Until the middle of the 1990s or so, effects of
inhomogeneities were not that important in observational cosmology, for
two reasons.  First, the uncertainty in the cosmological parameters was
large, comparable to (with respect to the effect on the distance as a
function of redshift) variation in the inhomogeneity parameter $\eta$.
Second, most observations were at low redshift, whereas $\eta$ is a
higher-order effect compared to the first- and second-order parameters
\hnull{} and \qnull{} 
\citep[see also \eqn{8} in][]{RKantowski98b}.  
The use of \tIas{} for the \mzr{} extended observations to higher
redshift.  Also, both this test as well as others had constrained the
cosmological parameters to a degree that the effect of $\eta$ could no
longer be ignored, which led to another revival of interest. 

In general, the effect of $\eta$ depends on angular scale: large angular
scales correspond to a fair sample of the \universe{} within the beam,
while this is not necessarily the case for small angular scales.  Since
supernovae have an angular scale of about $10^{-7}$ \arcsecw{}, which is
very small, one would perhaps expect to see effects of $\eta$ in the
\mzr{} for \tIas{}.  However, many independent investigations come to
the conclusion that $\eta\approx1$, not just on average, as is to be
expected, at least under certain assumptions 
\citep{SWeinberg76a}, but
also for each individual line of sight.  The reason for that is probably
that the 
\citet{YZeldovich64at}
model is incorrect in the sense that it is not a good approximation for
our \Universe{}: no-one doubts that the ZKDR distance is correct in a
\universe{} with a mass distribution well modelled by that on which the
idea of the ZKDR distance is based, but apparently that is not our
\Universe{}.  In other words, most of the matter is not outside the
beam, even for very narrow beams, but rather even such very narrow beams
fairly sample the \Universe{}.  Note that $\eta\approx1$ does not
necessarily imply that matter is distributed homogeneously within the
beam; it just implies that the distance as calculated from redshift is
approximately the same as if that were the case.  In reality, such a
beam will traverse voids with less than average density, but also
regions (corresponding to the filaments and sheets of large-scale
structure) with much higher than average density.  Although this
violates the assumptions on which the ZKDR distance is based,
nevertheless in practice such a mass distribution results in distance as
a function of redshift very close to the standard distance, \ie{}~that
optained by assuming that the \universe{}, at least with regard to light
propagation, is completely homogeneous.

\section*{\ack{}}

I thank both anonymous referees for thorough reviews of the first draft
and resulting helpful comments, which have been incorporated into the
final version.  I'm particularly grateful to Maria C. McEachern (John G.
Wolbach Library, Smithsonian Astrophysical Observatory) for supplying me
with copies of the original Russian articles by Zel'dovich, Dashevskii,
and Slysh.  This research has made use of NASA's Astrophysics Data
System Bibliographic Services.  The figure was made with help of the
\software{gral} software package written by Rainer Kayser.

\bibliographystyle{mnras}
\bibliography{abbrev,astro,astro_ph}

\end{document}